\documentclass[a4paper,fleqn,usenatbib]{mnras}
\pdfoutput=1

\usepackage[T1]{fontenc}
\usepackage{ae,aecompl}
\usepackage{hyperref}
\usepackage{graphicx}	% Including figure files
\usepackage{amsmath}	% Advanced maths commands
\usepackage{amssymb}	% Extra maths symbols
\usepackage[normalem]{ulem} %%%% FIXME ONCE DONE!
\usepackage{xspace}
\newcommand{\lumi}{\ensuremath{L_{\mathrm{radio}}/L_{\gamma}}\xspace}

\title[]{Relativistic protons in the Coma galaxy cluster: first gamma-ray constraints ever 
on turbulent reacceleration}

\author[G. Brunetti et al.]{
G. Brunetti,$^{1}$\thanks{E-mail: brunetti@ira.inaf.it}
S. Zimmer,$^{2}$\thanks{E-mail: zimmer@cern.ch}
F. Zandanel,$^{3}$\thanks{E-mail: f.zandanel@uva.nl}
\\
$^{1}$\it INAF-IRA, Via Gobetti 101, I-40129 Bologna, Italy\\
$^{2}$\it DPNC, University of Geneva, 24 Quai Ernest-Ansermet, CH-1211 Geneva, Switzerland\\
$^{3}$\it GRAPPA, University of Amsterdam, Science Park 904, 1098XH, Amsterdam, Netherlands
}

\date{Accepted XXX. Received YYY; in original form ZZZ}

\pubyear{2015}

\begin{document}
\label{firstpage}
\pagerange{\pageref{firstpage}--\pageref{lastpage}}
\maketitle

\begin{abstract}
The Fermi-LAT collaboration recently published deep upper limits to the
gamma-ray emission of the Coma cluster, a cluster that exhibits 
non-thermal activity and that hosts the prototype of giant radio halos.
In this paper we extend previous studies and
use a general formalism that combines particle
reacceleration by turbulence 
and the generation of secondary particles in the intracluster medium to 
constrain relativistic protons and their role for the origin of the 
radio halo.
Our findings significantly strengthen the conclusions of previous
studies and clearly disfavour pure secondary models for the halo. 
Indeed the new limits allow us to conclude that a pure hadronic origin of the 
radio halo would require magnetic fields that are too strong. For
instance $B_0 > 21 \mu$G is found in the cluster center assuming that the magnetic 
energy density scales with thermal density, to be compared with 
$B_0 \sim 4-5 \mu$G as inferred from Faraday Rotation Measures (RM) 
under the same assumption. 
Secondary particles can still generate the observed emission if they 
are reaccelerated. For the first time the deep gamma-ray limits allow us to
derive meaningful constraints if the halo is generated during phases 
of reacceleration of relativistic protons and their secondaries by 
cluster-scale turbulence. In this paper we explore a limited, but
relevant, range of parameter-space of reacceleration models.
Within this parameter space a fraction of model configurations 
is already ruled out by current gamma-ray limits, including the cases 
that assume weak magnetic fields in the cluster core, $B \leq 2-3 \mu$G.
Interestingly,
we also find that the flux predicted by a large fraction of model
configurations that assume a magnetic field consistent with Faraday RM 
is not far from the limits.
This suggests that a detection of cosmic-ray-induced gamma rays from the cluster might be possible 
in the near future, provided that the electrons generating the radio halo are secondaries 
reaccelerated
and the magnetic field in the cluster is consistent with that inferred from Faraday RM. 
\end{abstract}


%%%%%%%%%%%%%%%%%%%%%%%%%%%%%%%%%%%%%%%%%%%%%%%%%%

%%%%%%%%%%%%%%%%% BODY OF PAPER %%%%%%%%%%%%%%%%%%

\section{Introduction}

Clusters of galaxies form during the most violent events known in the Universe
by merging and accretion of smaller structures onto larger ones 
\cite[see][for a review]{Voit:2005aa}. This process
is accompanied by a release of energy of the order of the cluster gravitational binding energy of 
about $10^{62}-10^{64}$~erg. Some of this energy is dissipated via shock waves and turbulence 
through the intra-cluster medium (ICM) that can (re-)accelerate particles 
to relativistic energies \citep[see, e.g.][for a review]{Brunetti:2014aa}.
In fact, the presence of relativistic electrons in the ICM, as well 
as $\mu$G magnetic fields, is probed
by the detection of diffuse cluster-scale
synchrotron radio emission in several clusters 
of galaxies.
Mpc-scale diffuse synchrotron emission in clusters is classified in two
main categories: radio halos, roundish radio sources located in the
central regions, and radio relics, filamentary structures tracing
shock waves in the cluster outskirts \citep[][for a review]{Feretti:2012aa}
The origin of these synchrotron sources 
has been debated for decades now and, while some important steps have been made,
important ingredients in the scenario for the
origin of non-thermal phenomena in clusters 
remain poorly understood \citep[][for a review]{Brunetti:2014aa}.

\noindent
Giant radio halos are the most spectacular
evidence for non-thermal phenomena 
in galaxy clusters. The short cooling length of relativistic electrons
at synchrotron frequencies compared to the scale of these sources 
requires mechanisms of {\it in situ} acceleration or injection of the
emitting particles.
Cosmic-ray (CR) protons are accelerated by structure formation shocks 
and galaxy outflows in clusters, they can accumulate 
and are confined there for cosmological times and, therefore, 
can diffuse on Mpc volumes \citep[e.g.,][]{Volk1996,Berezinsky:1997aa}
For these reasons, a natural explanation for radio halos
was given by the so-called hadronic model \citep[{\it pure} secondary model
throughout this paper; e.g.,][]{Dennison:1980aa,Blasi1999,
Pfrommer2004,2008MNRAS.385.1211P,2010MNRAS.409..449P,
Keshet:2010aa,2011A&A...527A..99E}. 
In this scenario the observed diffuse radio emission in cluster
central regions is explained by secondary electrons that are
continuously generated by inelastic collisions between 
CR protons and thermal protons of the ICM. However the same inelastic 
collisions generate also gamma rays (from $\pi^0$ decay) whose non 
detection in these years \citep[e.g.,][]{2009A&A...502..437A,2009A&A...495...27A,
2010ApJ...710..634A,2012A&A...541A..99A,
Arlen:2012aa,Ackermann2010b,Huber:2013aa,Prokhorov:2014aa,Ahnen:2016aa}
limits the cluster CR content and in fact disfavors a pure hadronic origin 
of radio halos \citep[e.g.,][]{Brunetti:2009aa,Jeltema:2011aa,
Brunetti:2012aa,Zandanel:2014ab,Zandanel:2014aa,
Ackermann:2014aa,Ackermann:2016aa}.
While there are other complementary evidences that put tension on pure  
hadronic models, for example as inferred from cluster radio--thermal 
scaling relations and spectrum of radio halos \citep[][for review]{Brunetti:2014aa}, 
gamma-ray (and also neutrino; e.g., \citealp{Berezinsky:1997aa,2015A&A...578A..32Z})
observations are the only direct ways of constraining CRs in clusters.

Nowadays, the favored explanation for giant radio halos is given by 
the so-called turbulent re-acceleration model,
where particles are re-accelerated to emitting energies by turbulence 
generated during cluster mergers \citep{Brunetti:2001aa,Petrosian2001,Fujita:2003aa,Cassano2005,
Brunetti:2007aa,Donnert:2013aa,Beresnyak:2013aa,Miniati:2015aa,Brunetti:2016aa,
Pinzke:2017aa,2017arXiv170502341E};
a scenario that requires a complex hierarchy of mechanisms and plasma
collisionless effects operating in the ICM.
This scenario has the potential to fit 
well with a number of observations: from the radio halo -- merger connection, 
to spectral and spatial characteristics of the diffuse synchrotron emission.
However, one of the critical points in this scenario is the 
need of seed (relativistic) electrons for the re-acceleration \citep[e.g.,][]{Petrosian:2008aa}. 
A natural solution for this problem 
is again to assume that these seeds are secondary electrons coming from CR-ICM hadronic 
collisions \citep{Brunetti:2005aa,Brunetti:2011aa,Pinzke:2017aa}.
In this case, while the CR-proton content needed to explain radio halos 
is lower than in the pure hadronic scenario, one 
also expects some level of gamma-ray emission. 
Being able to test this level of emission would allow to
constrain the role of hadrons for the origin of the observed
non-thermal emission in galaxy clusters and also to constrain
fundamental physical parameters of the turbulence and of the
acceleration and diffusion of relativistic particles in the ICM.

The Coma cluster of galaxies hosts the best studied prototype of a 
giant radio halo \citep{Wilson:1970aa,Giovannini:1993aa}.
In this work we use the latest observations of the Coma 
cluster by the Large Area Telescope (LAT) on board of the Fermi
satellite - in particular the publicly available 
likelihood curves \citep{Ackermann:2016aa} - to constrain the role of CRp for the origin of
the halo, significantly extending previous studies by 
\citet{Brunetti:2012aa}, \cite{Zandanel:2014aa} and \citet{Pinzke:2017aa}.
Specifically we test a scenario based on pure hadronic models,
and a more complex scenario based on turbulent 
re-acceleration of secondary particles.
We use the surface brightness profile of the Coma giant
radio halo as measured by \citet{Brown:2011aa} at 350~MHz and the
synchrotron radio spectrum of the halo as priors for the modeling of the gamma-ray 
emission. The magnetic field is a very important ingredient in the
modeling as the ratio 
between gamma rays and the observed radio emission depends on the field 
strength and spatial distribution in the cluster.
We assume the magnetic field as free parameter 
while noting that Faraday Rotation Measures (RM) are available for 
Coma \citep{Bonafede:2010aa}, and we take 
these to be our baseline model.
We will show that we are now able to exclude a pure 
hadronic origin of the diffuse radio emission, 
unless extremely high magnetic fields - in clear contrast with 
RM - are assumed, therefore, extending previous
works in the same direction \citep{Brunetti:2012aa,Zandanel:2014ab,Zandanel:2014aa}. 
More importantly, we show for the first time that gamma-ray 
observations by Fermi-LAT allow now to start testing   
the turbulent re-acceleration scenario for the Coma giant radio 
halo, thereby opening up a new window into the study of
non-thermal phenomena in clusters with high-energy observations.
Interestingly, \cite{2015A&A...581A.126S} performed a re-analysis of the
Fermi-LAT data with an approach based on information
theory and found significant emission toward the direction of 
several galaxy clusters. However, due to the limitations of their analysis, 
it remains to be seen if these signals indeed belong to clusters or to point-sources. 
On a similar note, recently, \cite{2017ApJS..228....8B} reported the detection of cross-correlation
between the Fermi-LAT data and several galaxy cluster catalogues. Also in this case
it was not possible to disentangle between truly ICM-related diffuse emission and
point-sources. Nevertheless, these two works, together with the conclusions of our paper,
may point to a brighter future for gamma-ray observations of clusters.

This paper is organised as follows. In Sections~2 and 3, we introduce the 
physical scenario and the formalism adopted for our calculations.
In Section~4, we explain the method
used to obtain Fermi-LAT limits and the constrains on the combined
magnetic and CR properties from radio observations. 
We discuss our results on both pure hadronic
and the general re-acceleration scenarios in Section~5, 
and provide a discussion including the main limitations of the work 
in Section 6. Conclusions are given in Sction 7.

\noindent
In this paper we assume a $\Lambda$CDM cosmology with $H_0 = 70$km s$^{-1}$
Mpc$^{-1}$, $\Omega_{\Lambda}=0.7$ and $\Omega_m=0.3$.

\section{The physical scenario}

In this paper we explore the general scenario where CRp and their
secondaries are reaccelerated by ICM turbulence. 
This scenario has the potential to explain 
giant radio halos depending on the 
combination of turbulent-acceleraion rate and injection rate of
secondary particles \citep[e.g.,][]{Brunetti:2011aa,Pinzke:2017aa}. 
In addition, this scenario also produces gamma rays. However,
as turbulence in the ICM cannot reaccelerate electrons to TeV energies,
the gamma-ray emission is powered only by the process
of continuous injection of secondary particles due to proton-proton
collisions in the ICM ($\pi^0$-decay and inverse Compton (IC) from TeV secondary electrons)
and ultimately depends on the energy budget in the form of CRp in the
ICM at the epoch of observation.

\begin{figure*}
\includegraphics[width=\textwidth]{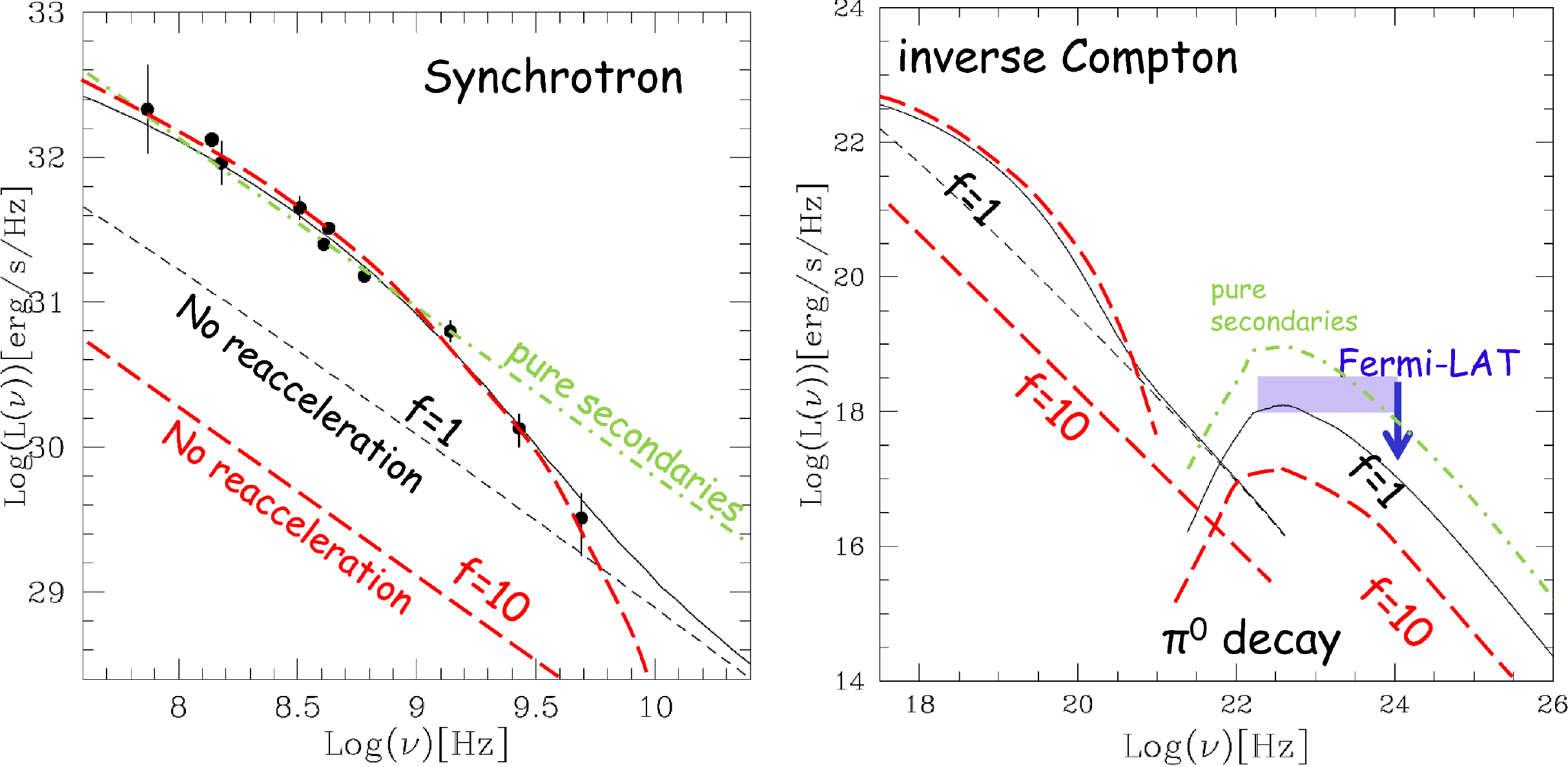}
\caption{An example of typical radio (left) and high-energy (right) spectrum 
predicted by reacceleration models that consider both primary and
secondary particles. The example is that of a Coma--like cluster
assuming a configuration of the magnetic field from Faraday RM and a reacceleration
rate that is suitable to match the steepening of the radio spectrum,
$\tau_\mathrm{acc}\sim 300$ Myr.
The cases $f=1$ (black) and $=10$ (red) are shown in the case of
reacceleration and normalised to the radio spectrum (i.e. with CRp number
tuned to match the radio spectrum during reacceleration). The
non-thermal emission from the two models assuming $\tau_{acc}^{-1}=0$ 
(no reacceleration, i.e. after turbulence is dissipated) is also
shown for comparison, note that
in this case the non-thermal luminosity scales about with $f^{-1}$.
The case of a pure secondary model anchored to the
observed radio emission at 1.4 GHz is also shown (green) together with
the coresponding gamma-ray spectrum that is expected due to $\pi^0$ decay.}
\label{fig1}
\end{figure*}

\noindent
For sake of clarity, in this Section we will briefly review the
expectations and main dependencies of the scenario. 
Calculations of (merger-induced) turbulent reacceleration 
of primary and secondary particles predict a complex spectral energy 
distribution (SED) of the non-thermal radiation that
extends from radio to gamma rays.
The SED is made of several components that have 
different evolutions in time; 
an example for the Coma cluster is given in Figure 1.

\noindent
In dynamically disturbed (merger) and turbulent systems, models predict
a {\it transient} component of the SED that is generated by turbulent
acceleration in the ICM. 
This {\it transient} component is expected to have a typical time-scale of about 
1-3 Gyr. During mergers particle acceleration by turbulence boosts up
the synchrotron and IC emission making the
cluster-scale non-thermal emission more luminous in the radio and hard 
X-ray bands (Fig. 1).
In fact the observed giant radio halos seems to be generated during 
this {\it transient} phase \citep[][for a review]{Brunetti:2014aa}.
The strength and spectral properties of the {\it transient} component
of the SED depend on the acceleration parameters and physical conditions in the ICM.

\noindent
In addition to the {\it transient} component, 
{\it long living} radio, IC and gamma-ray emission, sustained by the
continuous injection of high energy secondary CRe and by the decay of neutral 
pions, is expected to be common in clusters (Fig. 1).
The strength of this persistent component is proportional to
the energy density of the primary CRp and depends on their spatial
distribution in the ICM. Long living synchrotron emission can generate
under-luminous/off-state radio halos that may become visible in future
radio surveys \citep[e.g.,][]{Brown:2011ab,Cassano:2012aa}.

\noindent
The relative strength of the {\it transient} and {\it longer living}
component of the SED depends on the efficiency of particle 
acceleration and energy losses in the ICM.
Furthermore, in general,
this also depends on the ratio of primary and
secondary electrons that become available to the reacceleration process.
The ratio between the two populations is poorly known. 
Primary relativistic electrons can be accumulated and maintained
in galaxy clusters at energies of about 100 MeV \citep[][for a review]{Brunetti:2014aa}, 
so they may provide a suitable population of seeds
to reaccelerate
\citep[e.g.][]{Brunetti:2001aa,Cassano2005,Pinzke:2013aa,Pinzke:2017aa}.
For this reason it is convenient to define a parameter $f=1 +
{{ N_{e,pr} }\over{N_{e^{\pm}}}}$, where $N_{e,pr}/N_{e^{\pm}}$ is
the ratio of primary and secondary electrons that {\it become available} for
reacceleration; $f=1$ is the case where only CR protons (and their
secondaries) are present in the ICM. In Fig. 1 we show 
the differences in the SED 
that are caused by different assumptions for the ratio between primary 
and secondary electrons, $f=1$ and 10 respectively. 
Once the radio data points are used to constrain the amount of
reaccelerated electrons (other model parameters being fixed),
the gamma-ray luminosity declines with increasing $f$.
This is essentially because less
secondaries, and less CRp, are assumed in the model in order for it to
match the radio data.

If turbulent reacceleration does not play a role the scenario is that
of a pure secondary model where the spectrum of electrons and the SED
are simply governed by the injection of secondary
particles due to CRp-p collisions 
and by the radiative electron energy losses. 
Pure hadronic models have also been proposed for the origin of giant radio
halos \citep{Dennison:1980aa,Blasi1999,Pfrommer2004,Keshet:2010aa}. However,
in the last decade radio observations and their
combination with complementary constraints in the X-ray and gamma-ray
bands have put significant tension on this scenario \citep[][for a review]{Brunetti:2014aa}.
One reason for this tension originates from the non detection of galaxy
clusters in gamma rays as discussed in the Introduction.
This is shown in Figure 1 (green lines): the 
CRp energy budget that is necessary to match the synchrotron flux
of the radio halo is much larger than that in the
reacceleration models (other model parameters being the same) 
with the consequence to generate 
gamma rays in excess of observed limits 
from gamma-ray observatories, both space-borne and ground-based.
To circumvent this problem a larger value of the magnetic field must be 
assumed in the ICM. However it has been shown, for a number of nearby clusters,  
that the magnetic field required to explain radio halos without generating too many 
gamma rays is larger or in tension with independent estimates based on Faraday rotation
measures alone \citep[e.g.,][]{Jeltema:2011aa,Brunetti:2012aa} (see also
Sect. 5.1).

\section{Formalism adopted in the paper}

\noindent 
The aim of this Section is to present the essential information about the
formalism and the main 
assumptions used in our calculations.

\subsection{Time evolution of particles spectra}

We model the time evolution of the spectral energy 
distribution of electrons, $N_e^-$, and positrons, $N_e^+$,
with standard isotropic Fokker-Planck equations:
\begin{eqnarray}
{{\partial N_e^{\pm}(p,t)}\over{\partial t}}=
{{\partial }\over{\partial p}}
\Big[
N_e^{\pm}(p,t)\Big(
\left|{{dp}\over{dt}} \right|_{\rm rad} -
{1\over{p^2}}{{\partial }\over{\partial p}}(p^2 D_{\rm pp}^{\pm})
\nonumber\\
+ \left|{{dp}\over{dt}}\right|_{\rm i} \Big)\Big]
+ {{\partial^2 }\over{\partial p^2}}
\left[
D_{\rm pp}^{\pm} N_e^{\pm}(p,t) \right] \nonumber\\
+ Q_e^{\pm}[p,t;N_p(p,t)] \, ,
\label{elettroni}
\end{eqnarray}
\noindent
where $|dp/dt|$ marks radiative (r) and Coulomb (i) losses (Sect.~3.2),
$D_{pp}$ is the electron/positron diffusion coefficient in the
particles momentum
space due to the coupling with turbulent fluctuations (Sect.~3.3), 
and the term $Q_{e}^{\pm}$ accounts for the injection rate of secondary 
electrons and positrons due to p-p collisions in the ICM (Sect.~3.4).

\noindent
Similarly, the time evolution of the spectral energy
distribution of protons, $N_p$, is given by:
\begin{eqnarray}
{{\partial N_p(p,t)}\over{\partial t}}=
{{\partial }\over{\partial p}}
\Big[
N_p(p,t)\Big( \left|{{dp}\over{dt}}\right|_{\rm i}
-{1\over{p^2}}{{\partial }\over{\partial p}}(p^2 D_{\rm pp})
\Big)\Big]
\nonumber\\
+ {{\partial^2 }\over{\partial p^2}}
\left[ D_{\rm pp} N_p(p,t) \right] - {{N_p(p,t)}\over{\tau_{pp}(p)}} +
Q_p(p,t)
\, ,
\label{protoni}
\end{eqnarray}
\noindent
where $|dp/dt_i|$ marks Coulomb losses (Sect.~3.2),
$D_{pp}$ is the diffusion coefficient in the momentum
space of protons due to the coupling with turbulent 
modes (Sect.~3.3), $\tau_{pp}$ is the proton lifetime 
due to p-p collisions in the ICM (Sect.~3.4), and $Q_p$ is a source term.

\noindent
In our modelling 
the Fokker-Planck equations for protons, electrons and positrons are
coupled with
an equation that describes the time evolution of the spectrum of
turbulence, ${\cal W}(k,t)$.
For isotropic turbulence (see Sect.~3.3) the diffusion equation in the
k--space of the modes is given by:
\begin{eqnarray}
{{\partial {\cal W}(k,t) }\over{\partial t}}
=
{{\partial}\over{\partial k}}
\left(
k^2 D_{kk}
{{\partial}\over{\partial k}}
( {{ {\cal W}(k,t) }\over{k^2}} )
\right) + I(k,t) \nonumber\\
- \sum_i \Gamma_i (k,t) {\cal W}(k,t) \,\,\, ,
\label{modes_kinetic}
\end{eqnarray}
\noindent
where the relevant coefficients are the diffusion coefficient in the
k--space,
$D_{kk}$, the combination of the relevant 
damping terms, $\Gamma_i(k,t)$, and the
turbulence injection/driving term, $I(k,t)$ (see Sect.~3.3).

The four differential equations provide a fully coupled system:
the particle reacceleration rate is determined by the turbulent
properties, 
these properties are affected by the turbulent 
damping due to the same 
relativistic (and thermal) particles, and the injection rate and
spectrum of secondary 
electrons and positrons is determined by the proton spectrum which
(also) evolves with time. 

\subsection{Energy losses for electrons and protons}

The energy losses of relativistic electrons in the ICM are
dominated by ionization and Coulomb losses, at low energies 
($\leq$100-300 Myr), and by synchrotron and IC losses, 
at higher energies \citep[][for a review]{Brunetti:2014aa}.

The rate of Coulomb losses for electrons is dominated by 
the effect of CRe-e collisions in the ICM. 
For ultrarelativistic electrons the losses can be approximated as
(in cgs units):
\begin{equation}
\left| {{ d p }\over{d t}}\right|_{\rm i} 
\simeq 3.3 \times 10^{-29} n_{\rm th}
\left[1+ {{ {\rm ln}((p/m c)/{n_{\rm th}} ) }\over{
75 }} \right]
\label{e-e}
\end{equation}
\noindent
where $n_{\rm th}$ is the number density of the thermal plasma.
The rate of synchrotron and IC losses is (in cgs units):
\begin{equation}
\left| {{ d p }\over{d t}}\right|_{\rm rad}
\simeq 4.8 \times 10^{-4} p^2
\left[ \left( {{ B_{\mu G} }\over{
3.2}} \right)^2 %{{ \sin^2\theta}\over{2/3}}
+ (1+z)^4 \right]
\label{syn+ic}
\end{equation}
\noindent
where $B_{\mu G}$ is the magnetic field strength in
units of $\mu G$.
Eqs.\ref{e-e}--\ref{syn+ic} provide the coefficients in
Eq.~\ref{elettroni}.

For ultra-relativistic protons, in the energy range 10 GeV -- 100 TeV,
the main channel of energy losses 
in the ICM is provided by inelastic CRp-p collisions.
The lifetime of protons (in Eq.~ 2) due to CRp-p collisions is given by:
\begin{equation}
\tau_{pp}(p)=
{1\over{c \, n_{th} 
\sigma_{in} }}
\label{tpp}
\end{equation}
\noindent
where $\sigma_{in}$ is the total inelastic cross section. In this paper
we use the parameterization for the inelastic cross section given by
\cite{Kamae:2006aa,Kamae:2007aa} 
combining the contributions from the
non-diffractive, diffractive and resonant-excitation ($\Delta(1232)$
and res(1600)) parts of the 
cross section.

\noindent
In our calculations it is important also to follow correctly the
evolution
(acceleration and losses) of supra-thermal, trans-relativistic and
mildly 
relativistic protons. Indeed during reacceleration phases   
these particles might be reaccelerated and contribute significantly
to the injection
of secondary particles.
At these energies the particle energy
losses are dominated by Coulomb collisions due to CRp-p and CRp-e
interactions.
Following \citet{Petrosian:2015aa} the energy losses due to the
combined effect of CRp-p and CRp-e Coulomb interactions is (in cgs units):
\begin{eqnarray}
\left| {{ d p }\over{dt}} \right|_i \simeq
3.8\times 10^{-29} n_{th} 
\left(
1 - {{k_B T / (2 m_p c^2) }\over{ \sqrt{1 + (p/(m_pc))^2} -1  }}
\right) \nonumber\\
\times \big( {{\ln \Lambda }\over{38}} \big) \sum_{j=e,p} \left(
\int_0^{\sqrt{x_j}} \exp{(-y^2)} dy - \sqrt{x_j} \exp{(-x_j)} 
\right)
\label{pppe}
\end{eqnarray}
\noindent
where 
$\Lambda$ is the Coulomb logarithm, and $x_j = (E_p/k_B T)
(m_j/m_p)$, $j=e,p$ is for p-e and p-p interactions, respectively. 
Eq.~\ref{pppe} provides the coefficient for energy
losses in Eq.~\ref{protoni}.

\subsection{Turbulence and particle acceleration}

Particle acceleration in the ICM has been investigated in numerous
papers \citep{Schlickeiser:1987aa, Brunetti:2001aa,Brunetti:2004aa,Petrosian2001,Fujita:2003aa,
Fujita:2015aa,Cassano2005,Beresnyak:2013aa,
Brunetti:2016aa}, with particular focus on the effect of compressive turbulence induced
during cluster mergers
\citep{Brunetti:2007aa,Brunetti:2011aa,Miniati:2015aa,Brunetti:2016ab,Pinzke:2017aa}.
A commonly adopted mechanism is the Transit-Time-Damping (TTD)
\citep{Fisk:1976aa,Eilek:1979aa,Miller:1996aa,Schlickeiser:1998aa} with fast modes.
This mechanism is essentially driven by the resonance between the
magnetic moment of particles and the magnetic field gradients parallel to the 
field lines.
The mechanism induces a random walk in the particle momentum space
with diffusion coefficient $D_{pp}$ (Appendix A):
\begin{eqnarray}
D_{pp} \simeq
{{\pi^2}\over{2 c}}
{{c_s^2}\over{B_0^2}}
{{p^2}\over{\beta}}
\int_{{c_s}\over{\beta c}}^1
{\cal H}(1 - {{c_s}\over{\beta c}})
{{1 - \mu^2}\over{\mu}} d \mu
\left[
1 - \big( {{c_s}\over{\mu \beta c}} \big)^2
\right]^2 \nonumber\\
\times \int_{k_0}^{k_{cut}} dk k W_B
\label{dppnew}
\end{eqnarray}
\noindent
where $\mu$ is the cosine of the particle pitch angle, 
$W_B$ is the spectrum of turbulent magnetic fluctuations, $k_0$ and $k_{cut}$ 
are the injection and cut-off wavenumbers respectively, and ${\cal H}$ is the
Heaviside function.
Eq.~\ref{dppnew} provides the diffusion coefficient in the particles
momentum space that is used in Eqs.\ref{elettroni}--\ref{protoni}.
For ultra-relativistic particles
the resulting particles acceleration time, $\tau_\mathrm{acc} = p^2/(4
D_{pp})$, is independent from the particle momentum.

More specifically in this paper we obtain 
the spectrum of magnetic turbulence in Eq.~\ref{dppnew}, $W_B$, 
from Eq.~\ref{modes_kinetic} assuming 
$|B_k|^2 \sim 16 \pi \beta_\mathrm{pl} W$
\citep[see][]{Brunetti:2007aa}, $\beta_\mathrm{pl}$ is the
plasma beta.
In first approximation, this spectrum is expected to be a power-law induced by
wave-wave cascade
down to a cut-off scale, $k_{cut}$, where the turbulent-cascading 
time becomes comparable to or longer than the damping time,
$\Gamma^{-1}$. 
The cut-off scale and the amount of turbulent energy
(electric$/$magnetic field
fluctuations)
that is associated with the scales
near the cut-off scale
set the rate of acceleration. In this paper we adopt a Kraichnan
treatment to describe 
the wave-wave cascading of compressible MHD waves;
this gives \citep[see, e.g.,][]{Brunetti:2016ab}:
\begin{equation}
k_{cut} \simeq {{81}\over{4}}
\left(
{{\delta V^2}\over{c_s}}
\right)^2
{{ k_o }\over{ ( \sum_{\alpha} \langle 
\Gamma_{\alpha} \rangle k^{-1} )^2}}
\label{kcut}
\end{equation}
\noindent
where the velocity of large-scale eddies is $\delta V \sim \sqrt{
W(k_o)k_o}$, $k_0^{-1}$ is the injection scale and 
$<..>$ marks pitch-angle averaging.
The damping rate $\sum \langle \Gamma \rangle$ in Eq.~\ref{kcut} is
contributed also by the collisionless 
coupling with the relativistic particles
\citep{Brunetti:2007aa,Brunetti:2016ab} and consequently
the cut-off scale, acceleration rate and the spectrum of particles are derived
self-consistently from the solution of the four coupled equations in Sect. 3.1.
Two scenarios can be considered
to estimate the damping of fast modes in the ICM 
\citep[e.g.,][]{Brunetti:2016ab}.
(i) One possibility is that 
the interaction between turbulent modes with both thermal and CRs
is fully collisionless.
This happens when the collision frequency between particles in the ICM is
$\omega_{ii} < \omega = k c_{s}$; for example this is the case where ion-ion
collisions in the thermal ICM are due to Coulomb collisions.
(ii) The other possibility is that 
only the interaction between turbulence and CRs is collisionless.
This scenario is motivated by the fact that 
the ICM is a {\it weakly collisional} high-beta plasma that is unstable
to several instabilities that can increase the collision frequencies
in the thermal 
plasma and induce a collective interaction of individual ions with the
rest of the plasma \citep[see e.g.,][]{Schekochihin:2006aa,
Brunetti:2011ab,Santos:2014aa, Kunz:2014aa, Santos:2017aa}.
\noindent
In this paper we adopt the most conservative scenario that is based on
a fully
collisionless interaction between turbulence and both thermal
particles and CRs (case i).
In this case the damping of compressive modes is dominated by TTD with
thermal particles \citep{Brunetti:2007aa} which also allows to greatly reduce the
degree of coupling
between Eqs. 1--3 and to simplify calculations.
For large beta-plasma, $\beta_\mathrm{pl}$, 
the dominant damping rate is due to thermal
electrons $\Gamma_e \sim
c_{s} k \sqrt{3 \pi (m_e/m_p) / 20 \mu^2} \exp( -5 (m_e/m_p)/3
\mu^2)(1 -\mu^2)$, leading 
to (from Eq.~\ref{kcut}) $k_{cut} \sim 10^4 k_0 M_0^4$, where $M_0 = \delta V / c_s$ 
is the turbulent Mach number.

\subsection{Injection of Secondary Electrons}

The decay chain that we consider for the injection
of secondary particles due to CRp-p
collisions in Eq.~1 is \citep[e.g.][]{Blasi1999}:
$$p+p \to \pi^0 + \pi^+ + \pi^- + \rm{anything}$$
$$\pi^0 \to \gamma \gamma$$
$$\pi^\pm \to \mu^\pm + \nu_\mu(\bar{\nu}_\mu) \,\,\,\,\, , \,\,\,\,\,
\mu^\pm\to e^\pm + \bar{\nu}_\mu({\nu}_\mu) + \nu_e(\bar{\nu}_e).$$
\noindent
that is a threshold reaction that requires CRp with kinetic
energy larger than $T_p \approx 300$ MeV. 
The injection rate of charged and neutral pions is given by:
\begin{equation}
Q_{\pi}^{\pm,0}(E,t)= n_{th} c 
\int_{p_{*}} dp N_p(p,t) \beta_p 
{{ d \sigma^{\pm, 0} }\over{ d E }} (E_{\pi},E_p) 
\label{q_pi}
\end{equation}
\noindent
where $d \sigma^{\pm, 0}/d E$ is the differential inclusive cross 
section for the production of charged and neutral pions.
A practical and useful approach that we adopt in this paper 
to describe the pion spectrum both in the high energy and low energy 
regimes is based on the combination of the isobaric and
scaling model \citep{Dermer:1986aa,Moskalenko:1998aa}.
Specifically, in this paper 
we consider four energy ranges to obtain a precise
description 
of the differential cross section from low to high energies:
\begin{itemize}
\item{for $E_p \leq 3$ GeV we use the isobaric model, specifically we
use Eqs.~23-28 in \citet{Brunetti:2005aa};}
\item{for $3 \leq E_p/{\mathrm{GeV}} \leq 7$, we use a combination of
the isobaric model with the parametrization given in
\citet{Blattnig:2000aa}, specifically their Eqs.24,26,28;}
\item{for $7 \leq E_p/{\mathrm{GeV}} \leq 40$, we use a combination of the aforementioned Blattnig-parametrization and the scaling model from \citet{Kelner:2006aa} based on QGSJET simulations, Eqs. 6--8;}
\item{finally, for $E_p > 40$ GeV we use the aforementioned scaling model.}
\end{itemize}

\noindent
The decay of $\pi^0$--decay generates gamma rays with spectrum:
\begin{equation}
Q_{\gamma}(E_{\gamma},t)= 2 \int_{E_{min}}^{E_{p,max}}
{{Q_{\pi}^0(E_{\pi},t) }\over{\sqrt{E_{\pi}^2 - m_{\pi}^2 c^4}}}
dE_{\pi}
\label{gammarays}
\end{equation}
\noindent
where $E_{min}=E_{\gamma} + m_{\pi}^2 c^4/(4 E_{\gamma})$.
The injection rate of relativistic electrons/positrons is given by:
\begin{eqnarray}
Q^{\pm}_{e}(p,t)=
c \int_{E_{\pi}}
Q^{\pm}_{\pi}(E_{\pi^{\pm}},t) dE_{\pi} \int dE_{\mu} \times \nonumber\\
F_{e^{\pm}}(E_{\pi},E_{\mu},E_e) F_{\mu}(E_{\mu},E_{\pi}),
\label{qepm1}
\end{eqnarray}
\noindent
where $F_e^{\pm}(E_e,E_\mu,E_\pi)$ is the spectrum of
electrons and positrons from the
decay of a muon of energy $E_\mu$ produced in the decay of a pion with
energy $E_\pi$, and $F_{\mu}(E_{\mu},E_{\pi})$ is the muon spectrum
generated by the decay of a pion of energy $E_{\pi}$.
In this paper we follow the approach in \citet{Brunetti:2005aa} to
calculate Eq.~\ref{qepm1}, in this case the injection rate of
secondary electrons and positrons is:
\begin{eqnarray}
Q^{\pm}_{e}(p,t) =
{{ 8 \beta_{\mu}^{\prime} m_{\pi}^2 n_{th} c^2}\over{ m_{\pi}^2 - 
m_{\mu}^2 }}
\int_{E_{\rm min}}
\int_{p_{*} } 
{{dE_{\pi} dp}\over{ E_{\pi} {\bar{\beta}}_{\mu} }}
\beta_p N_p(p,t) \nonumber\\
\times {{ d \sigma^{\pm}}\over{d E}} (E_{\pi}, E_p) 
F_e(E_e,E_{\pi}) \, ,
\label{qepm2}
\end{eqnarray}
\noindent
where $F_e(E_e,E_{\pi})$ is given in \citet{Brunetti:2005aa} (their
Eqs.~36-37), ${\bar{\beta}}_{\mu}= \sqrt{
1 - m_{\mu}^2/\bar{E}_{\mu}^2}$, ${\bar{E}_{\mu}} =
1/2 E_{\pi} (m_{\pi}^2 - m_{\mu}^2)/(\beta_{\mu}^{\prime} m_{\pi}^2)$,
and $\beta_{\mu}^{\prime} \simeq 0.2714$.
Eq.~ \ref{qepm2} is the injection spectrum used in Eq.~ \ref{elettroni}.

\subsection{A note on pure hadronic models}

If turbulence is not present (or negligible acceleration) our formalism 
describes a pure hadronic model where the spectrum of electrons is
simply determined by the balance between the injection rate due to CRp-p
collisions and the cooling due to radiative and Coulomb losses.

Under these conditions, if the proton spectrum evolves slowly (on time-scales that are 
much longer than those of electron 
radiative and Coulomb losses) after a few electron cooling times, 
the spectrum of electrons reaches quasi-stationary conditions.
From Eq.~\ref{elettroni}, with $\partial N/\partial t =0$ and
$D_{pp} = 0$, one then has \citep[e.g.,][]{Dolag:2000aa}:
\begin{equation}
N_e^{\pm}(p; t)=
{ 1 \over { \sum_{r,i} \big| {{dp }\over{dt}} \big| }}
\int_p Q^{\pm}_{e} (p ; t) p dt \, .
\label{statpure}
\end{equation}

We note indeed that the timescale of CRp-p collisions of relativistic protons 
in the ICM is much longer than the radiative (and Coulomb) lifetime
of GeV
electrons \citep[see, e.g.][for reviews]{Blasi:2007aa,Brunetti:2014aa}, 
and consequently the use of Eq.~\ref{statpure} is fully justified to
model the spectrum of radio-emitting electrons.

\section{Method}

The main goal of this paper is to constrain the role of relativistic
protons for the origin of the observed radio emission in the Coma
cluster. To do that we model the observed spectrum and brightness
distribution of the Coma radio halo assuming different CR scenarios and
compare the expected gamma-ray emission with the
upper limits derived from the Fermi-LAT.

\subsection{Constraints from the radio emission}

The recent WSRT observations of the Coma halo at 350 MHz \citep{Brown:2011aa} have 
reached an unprecedented sensitivity to low surface
brightness emission on large scales and allowed to derive solid
measurements of the brightness of the halo up to large,
$\sim 1$ Mpc, distance from the center \citep{Brunetti:2012aa,Brunetti:2013aa,Ade:2013ab,Zandanel:2014aa}.
In Figure 2 we show the integrated flux of the Coma halo at 350
MHz as a function of distance.\footnote{Obtained by \citet{Brunetti:2013aa} using the azimuthally
averaged profile of the halo and excluding the west quadrant due to the contamination 
from the radio galaxy NGC4869. Error bars in Figure 2 include both statistical and 
systematical uncertainties.}% [FZ: what are the systematics that are considered?]}.}
This figure shows that the halo is clearly detected up to 1 Mpc 
distance from the center and that the majority of the emission of the 
halo is produced in the range of distances 0.4-1.0~Mpc.
We note that the profile derived from these WSRT observations 
is consistent with other profiles at 330 MHz \citep{Govoni:2001} and
140 MHz \citep{Pizzo:2010aa}, at least within the aperture radius 
where these oldest observations are sensitive enough. 
On the other hand it is inconsistent with the profile obtained by
\citet{Deiss:1997aa} using Effelsberg observations at 1.4 GHz, possibly 
suggesting systematics or flux calibration errors in the 
observations at 1.4 GHz \citep[see eg.,][]{Pinzke:2017aa}.

\begin{figure}
\includegraphics[width=0.45\textwidth]{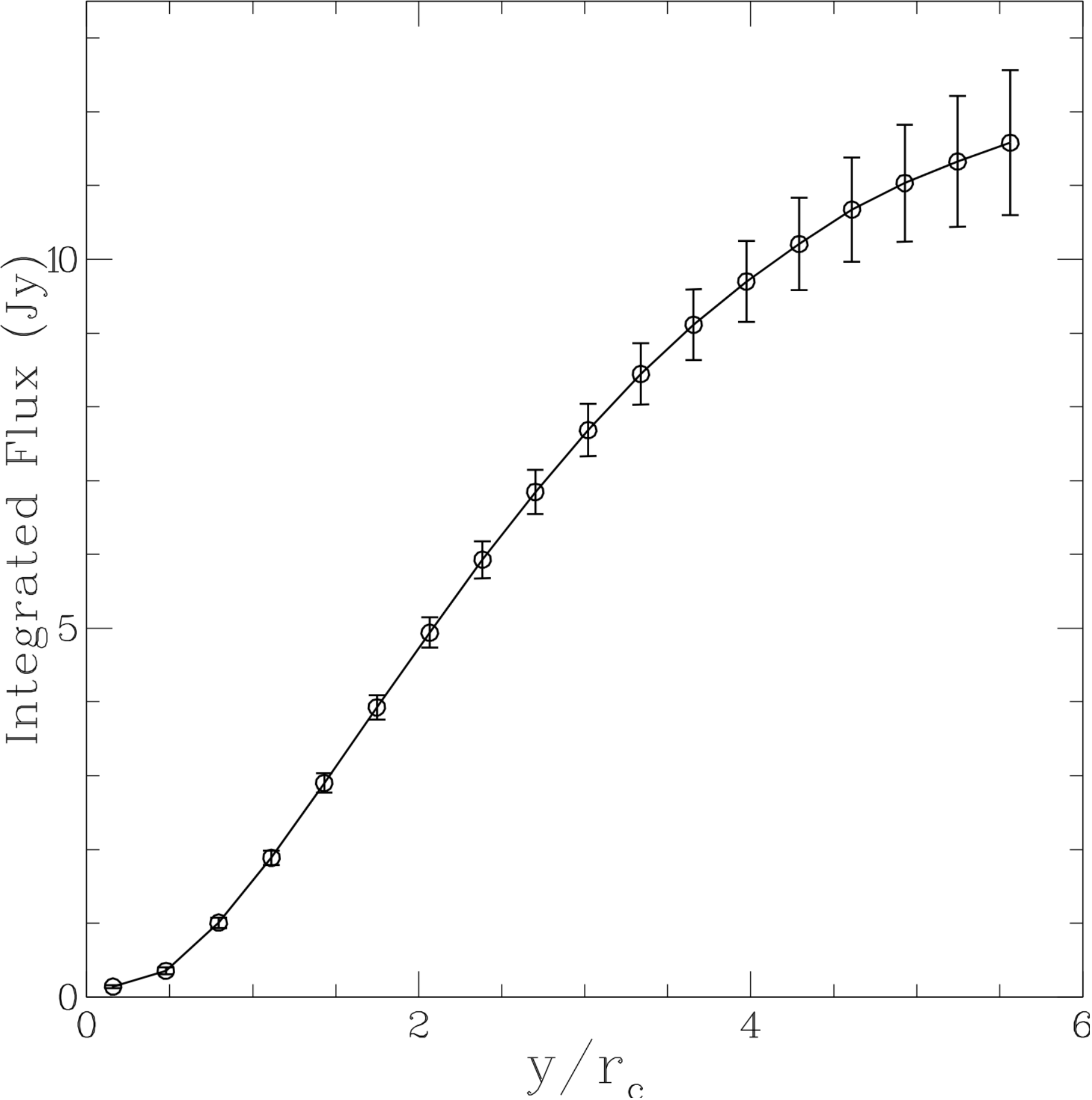}
\caption{The radio flux of the Coma cluster measured at 325 MHz as a
function of the projected radial distance (in unit of core radius; $R_{500}\simeq 5 r_c$). 
Note that the integrated profile is obtained excluding
the quadrant containing NGC 4869 and multiplying by $4/3$ the flux
measured within the three quadrants (see \citealp{Brunetti:2013aa} for
details).}
\label{fig.2}
\end{figure}

\noindent
If the radio halo is due to secondary particles,
the observed profile constrains spatial distribution 
and energy density of CRp for a given configuration 
of the magnetic field and model of the CRp spectrum. 
The additional needed ingredient is the the spatial distribution 
of the thermal target protons obtained from the
X-ray observations; we adopt the beta-model profile 
derived from ROSAT PSPC observations \citep{Briel:1992aa} with $r_c = 290$ kpc, $\beta =
0.75$ and central density $n_{\mathrm{ICM}}(0)= 3.42 \times 10^{-3}$cm$^{-3}$, $r_c$ being 
the core radius of the cluster.
When necessary, in order to derive the ratio of CRp to thermal energy densities, we also use the temperature 
profile of the Coma cluster taken from XMM and Suzaku observations \citep{Snowden:2008aa,Wik:2009aa}.

\noindent
The magnetic field in the Coma cluster has been constrained from RM \citep{Bonafede:2010aa,Bonafede:2013aa}. 
Following these studies, in our paper we assume a scaling of the magnetic field strength with cluster thermal 
density in the general form:
\begin{equation}
B(r) = B_0 \left(
{{ n_{\mathrm{ICM}}(r) }\over{n_{\mathrm{ICM}}(0) }}
\right)^{\eta_B}
\label{B}
\end{equation}
\noindent
where $B_0$ and $\eta_B$ are free parameters in our calculations. 
As a reference, $B_0 = 4.7 \mu$G and $\eta_B =0.5$
provides the best fit to RM data \citep{Bonafede:2010aa}. 

In addition to the brightness profile, the 
other observable that we use to constrain model parameters is 
the spectrum of the halo that is measured across more than
two decades in frequency range \citep{Thierbach:2003aa,Pizzo:2010aa}.
The radio spectrum constrains the spectrum of electrons and 
if the halo is generated by secondary particles 
it can be used to put constraints on the spectrum of CRp. 
This link however is not straightforward.
For example the spectral slope
of the radio halo in reacceleration models is more sensitive to
the reacceleration 
parameters rather than to the spectrum of the primary CRp.
For this reason we will discuss the constraints from the spectrum of
the halo separately in the case of pure hadronic models (Sects. 5.1) and 
in the general scenario (Sect. 5.2).

\subsection{Gamma-ray limits from Fermi-LAT}

Deriving meaningful limits to the gamma-ray luminosity and CRp energy content
of the Coma cluster is not straightforward because the gamma-ray flux 
and its brightness distribution depend on the spatial 
distribution of CRp and on their spectrum.
The most convenient approach is thus to generate models of CRp
(spectrum and spatial distribution) that explain the
radio properties of the halo, calculate the expected gamma-ray emission model
(spectrum and brightness distribution), and
determine if a specific cluster model is in conflict with Fermi-LAT data.
While the models and results are discussed in Sect.~5.1--5.2, here we
describe the general procedure that is adopted in the paper to obtain the 
corresponding gamma-ray limits (see also Section 6.1 for a discussion 
regarding the expected sensitivity of LAT observations with respect to these models).

We make use of the latest published likelihood curves in \cite{Ackermann:2016aa}
which investigated the Coma cluster using the latest data release from
the Fermi Large Area Space telescope ("Pass 8"). In contrast to
previous analyses \citep{Ackermann2010b,Zandanel:2014aa,Ackermann:2014aa}, 
this analysis reported a faint residual excess, however below the
threshold of claiming a detection. In addition to considering a set of
baseline spatial templates, the analysis presented in \citet{Ackermann:2016aa} 
also provided a set of likelihood curves per energy bin for 
disks of varying radii around the center of 
the Coma cluster ($\alpha_{2000}=194.95$, $\delta_{2000}=27.98$).

We assume that any CR-induced gamma-ray template 
can be modeled as disk of unknown radius $r$ (Appendix B),  
and devised the following approach
to determine whether or not the predicted gamma-ray emission of a specific CR model is in conflict with
the data presented in \citet{Ackermann:2016aa}.
For a given physical model $M$, based on the radio constraints (Sect.~4.1, Sect.~5.1--5.2), 
we calculate the predicted gamma-ray spectrum and the predicted
surface brightness profile. We then use {\tt{gtsrcmaps}} to fold this
physical model with the instrument response function (IRF)
corresponding to the Pass 8 event analysis obtaining for each energy bin $E_i$
a 2D intensity map.\footnote{In order to have
a fair comparison, we use the same version of the Fermi Science Tools
and the associated P8R2\_SOURCE\_V6 IRFs, separately for front- and
back-entering events \citep[see][for details on the
analysis]{Ackermann:2016aa}.}
We repeat the same procedure for
isotropic disks $D_{j}$ with radii ($0.1\times\theta_{200} -
1.0\times\theta_{200}$) with $\theta_{200}=1.^{\circ}23$ being the
subtended angle on the sky corresponding to the cluster virial radius
($\simeq$2.0~Mpc at $z=0.023$). In the next step we compare the IRF-folded model $M$ with each disk $D_{j}$. 
We find the closest match between $M$ and $D_j$ by taking into account the spectral shape (using 
the spectrum as weighting factor) and extract the tabulated ``bin-by-bin likelihood" 
$\mathcal{L}_{i}(\mu_{i}|D_{i})$ for the corresponding disk $D_j$.\footnote{Note that we tested 
different comparison operators, such as the average over all energy bins, a power-law spectrum and the physical 
spectrum provided by the model $M$, and found only marginal differences in 
the resultant upper limits on the gamma-ray flux.}
From our cluster model $M$ we also have the predicted model counts $\mu_{i}$ 
which are determined up to an overall 
normalization. By combining each bin-wise likelihood, we can form a joint (profile) likelihood function 
$\mathcal{L}$ \citep[Eq.~ 2 in][]{Ackermann:2016aa}:
\begin{equation}
\mathcal{L}(\mu,|D) = \prod_{i} \mathcal{L}_{i}(\mu_{i}|D_{i}) \, ,
\end{equation}
and calculate the 95\% C.L. flux upper limits for any model $M$ by finding 
the value for $\mu$ for which the difference in the log-likelihood with 
respect to the best-fit value of the alternative hypothesis (including 
the cluster model) equals 2.71.\footnote{Comparing our equation with that 
in \citet{Ackermann:2016aa}, it should be noted that nuisance parameters $\theta$ which have been profiled 
over $\hat{\theta}$ are omitted for simplicity.}

\section{Results}

In this Section we present the results from the comparison of expectations from
different CR models and gamma-ray limits. 

\subsection{Pure hadronic models}

In the case of pure secondary models, turbulent reacceleration does not 
play a role (Sects.~2 and 3.1). This is a special case in the general 
scenario described in Sect. 2.

\noindent
In this case the ratio between 
the synchrotron luminosity from secondary electrons and the gamma-ray
luminosity from $\pi^0$ decay is:
\begin{equation}
{{L_{radio}}\over{L_{\gamma}}} \propto \langle {{B^{\alpha +1}}\over{ B^2 +
B_{cmb}^2}} \rangle
\label{ratioradiogamma}
\end{equation}
\noindent
where $\alpha$ is the synchrotron spectral index and
$\langle .. \rangle$ indicates a volume-averaged quantity that is weighted for the spatial distribution of CRp.
Therefore, the combination of gamma-ray limits
and radio observations constrains a lower boundary of $B$ values 
\citep[e.g.,][]{Blasi:2007aa,Jeltema:2011aa,Arlen:2012aa,Brunetti:2012aa,Ahnen:2016aa}.
More specifically, by 
assuming a magnetic field configuration from
Eq.~\ref{B}, the gamma-ray limits combined with the
radio spectrum of the halo allow to
determine the minimum central magnetic field
in the cluster, $B_0$, for a given value of $\eta_B$ and for a given spatial 
distribution and spectrum of CRp. 
Using this procedure, \citet{Brunetti:2012aa} concluded that the minimum magnetic field implied 
by limits in \cite{Ackermann2010b} is larger than what previously has been estimated from Faraday RM.
\noindent
Fermi-LAT upper limits from \citet{Ackermann:2016aa} significantly
improve constraints from previous studies, such as
\citet{Ackermann2010b} and \citet{Ackermann:2014aa}.
Thus we repeat the analysis carried out by \citet{Brunetti:2012aa} using the
new limits. However, in addition, we also follow the more complex approach
described in Sect.~4.2. 
Using a grid of values $(B_0,\eta)$, 
we generate pure hadronic models that are anchored to the radio spectrum
of the Coma halo and its brightness profile 
and re-evaluate the Fermi-LAT limits for each model.
The comparison between these limits and the gamma-ray
flux produced in each case allows us to accept/rule out the
corresponding models thus deriving corresponding lower limits to $B_0$ 
(for a given $\eta_B$). 

\noindent
In the case of pure hadronic models a direct connection exists between
the spectrum of CRp and that of the radio halo.
Assuming a power-law in the form $N \propto p^{-\delta}$, it is
$\delta = 2 \alpha + \Delta$ \citep[e.g.,][]{Kamae:2006aa}, 
where $\Delta \sim 0.05$ (for typical spectra of radio halos)
accounts for the Log-scaling of the proton-proton cross section.
If we restrict ourselves to frequencies $\leq 1.4$ GHz (at higher
frquencies a spectral break is observed), the spectrum of the radio halo 
is well fitted using a power-law with slope
$\alpha = 1.22 \pm 0.04$. The data however show significant
scattering that is likely due to the fact that flux measurements are
obtained from observations with different sensitivities and using
different observational approaches (e.g., in the
subtraction of discrete sources, and in the area used to extract the
flux of the halo).
The spectral slope, however, does not seem to be affected very much 
by systematics. For example a slope $\alpha = 1.17 \pm 0.02$ is
obtained using only the three data-points
at 150, 330 and 1.4 GHz, where fluxes have been measured within the 
same aperture radius of about $530$ kpc ($y/r_c \sim 1.8$ in Fig.~2;
\citealp{Brunetti:2013aa}).
The observed range of spectral slopes of the halo 
selects a best fit value for the spectrum of CRp
$\delta \simeq 2.45$ and a 1$\sigma$ range of values $\delta = 2.35-2.57$ 
including the systematics due to different apertures used to measure the 
flux of the halo.
We adopt $\delta =2.45$ as reference value. Steeper (flatter) spectra provide 
stronger (weaker) limits, i.e. larger (smaller) magnetic fields are
constrained 
for larger (smaller) 
values of $\delta$ \citep{Brunetti:2009aa,Brunetti:2012aa,Zandanel:2014aa}.
More specifically, in order to have a realistic spectrum of CRp, we assume
a scenario where 
CRp are continuously injected for a few Gyrs with a power-law
spectrum and are subject to 
the relevant mechanisms of energy losses described in Sect. 3.2 
(we assume a typical {\it average} number density of the ICM
$n_{th}=10^{-3}$cm$^{-3}$ that would also account for the fact that in
a time-period $>$ Gyr CRp can efficiently diffuse/be advected in the
cluster volume).
The main outcome of this scenario is that the resulting spectrum of CRp
flattens at kinetic energies $<$~GeV due to Coulomb losses in the ICM.

Figure 3 (left) shows the minimum magnetic field that is allowed by the
combination of radio and gamma--ray data as a function of $\eta_B$.
As expected the lower limits to
magnetic field strengths are very large, significantly 
larger than those obtained by \citet{Brunetti:2012aa}.
In Figure 3, we also show the measurements of $(B_0,\, \eta_B)$ ($1-3\sigma$) from 
Faraday RM by \cite{Bonafede:2010aa}.
The latter constraints are obtained combining measurements for
7 radio galaxies (background and cluster sources) that cover an area 
comparable to that of the radio halo. 
In addition, 
the shaded region in Fig. 3 (left) represents the $3\sigma$ allowed
region of parameters using only the 5 background radio
galaxies reported in \citet{Bonafede:2010aa}. 
In principle the use of background sources 
provides a more solid approach because it may reduce the presence of
biases in the Faraday RM induced by the effects due 
to the interaction between the relativistic plasma in radio galaxies
and the local ICM \citep{Carilli:2002aa,Rudnick:2003,Bonafede:2010aa},
however the absence of background radio galaxies 
projected along the core of the Coma cluster makes RM constraints 
less stringent.

\begin{figure*}
\includegraphics[width=0.45\textwidth]{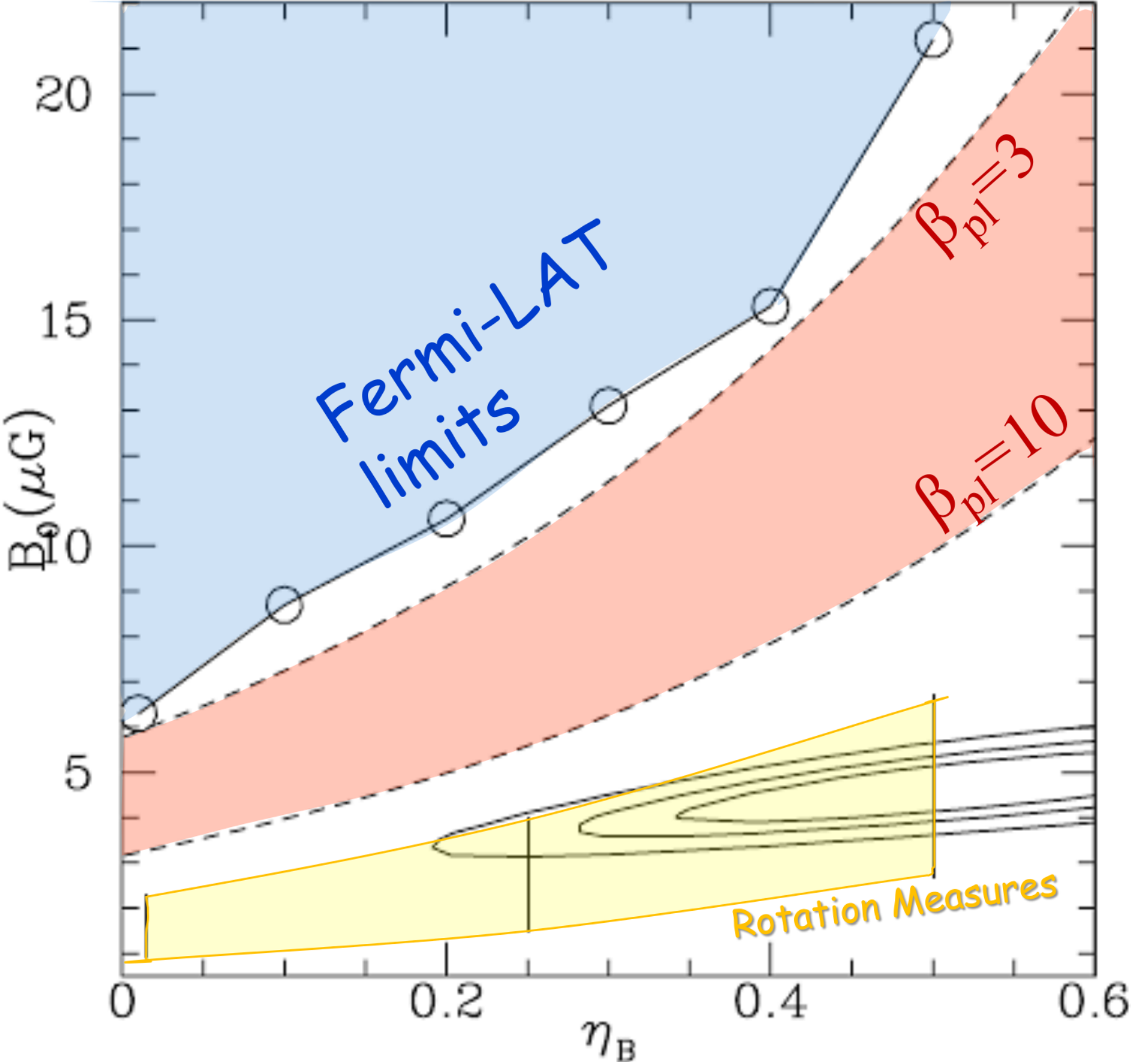}
\includegraphics[width=0.445\textwidth]{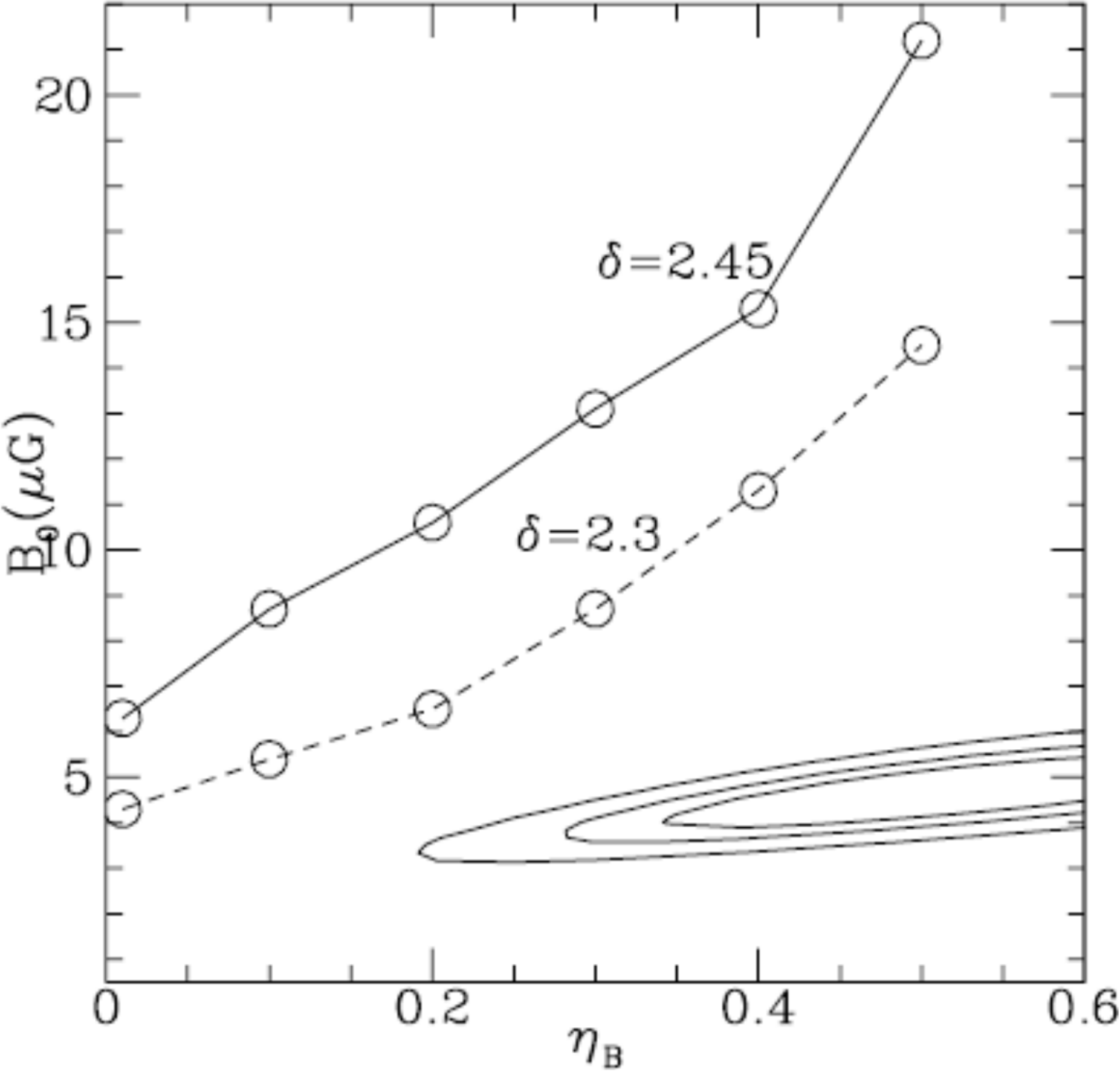}
\caption{(left) The area constrained by Fermi-LAT limits in the
$B_0-\eta_B$ diagram (blue area) assuming a pure hadronic models for
the Coma radio halo and $\delta=2.45$. This region is compared with the 1, 2, 3$\sigma$
constraints from Faraday RM (solid lines from \citealp{Bonafede:2010aa})
and with the region constrained from the same Faraday RM using only
background radio galaxies (yellow region; see main text for details).
The region spanned by $B_0$ and $\eta_B$ in the case of a $\beta_\mathrm{pl}=3$ to
10 at a distance of 2.5-3$r_c$ is also reported for comparison (orange
region).
(right) The same comparison between magnetic field lower limits and
constraints from Faraday RM assuming $\delta=2.45$ and 2.3.}
\label{fig.4}
\end{figure*}

\noindent
The discrepancy between minimum B-fields allowed for a 
pure secondary origin of the halo and that from RM is
very large, with the magnetic field energy density in the
cluster being $>$15-20 times larger than that constrained
using RM.
\noindent
An obvious consequence is that the magnetic field in the
ICM would be dynamically important in the case of a pure hadronic
origin of the halo. This is particularly true in the 
external, $r \geq 2-2.5 r_c$ (that is $r \geq 0.4-0.5 R_{500}$), 
regions where most of the cluster thermal
energy budget is contained and where most of the radio halo emission is
generated (Fig. 2).
In order to further explain this point, in Fig. 3
(left) we also report the region of parameters corresponding to a  
beta plasma (ratio of thermal and magnetic pressure, $\beta_\mathrm{pl} 
= 8 \pi P_{\mathrm{ICM}}/B^2$) between 3 and 10
at a reference distance 2.5-3 $r_c$ ($\simeq$730-875 kpc, 
i.e., $\simeq 0.5-0.6\times R_{500}$).
Limits on $B_0$ ($\eta_B$) from our combined radio --
Fermi-LAT analysis select $\beta_\mathrm{pl} < 2-3$ at these distances 
independently of the value of $\eta_B$ and imply 
an important (even dominant) contribution to cluster dynamics 
from the magnetic field
pressure. This is in clear
tension with several independent theoretical arguments and
observational constraints \citep[e.g.,][and references therein]{Miniati:2015ab}. 

\noindent
In Fig.3 (right) we show the effect of different
spectra of the CRp. Flatter spectra make constraints slightly less
stringent. This is simply because GeV photons produced by the decay of
neutral pions probe CRp with energies that are about 5-10 times smaller than 
those of the CRp which generate the secondary electrons emitting at 300 MHz.
However, if the spectrum of CRp has a slope $\delta > 2.35$, as
constrained by radio observations, our conclusions remain essentially
unaffected. 

\noindent
Our calculations assume a power-law spectrum of the primary CRp (subject to modifications
induced by Coulomb and CRp--p losses). Propagation effects in the ICM
might induce additional modifications in the spectrum of CRp that may
change the radio to gamma-ray luminosity ratio, \lumi, with respect to our
calculations. An unfavorable situation is when 
confinement of CRp in Mpc$^3$ volumes is inefficient for CRp energies
$\geq 10$ GeV.\footnote{However, note that this is unlikely as it is currently thought 
that CRp are confined galaxy clusters up to much greater energies
\citep[see, e.g.,][for reviews]{Blasi:2007aa,Brunetti:2014aa}}
In this case the spectrum of CRp at lower energies
may be flatter than that constrained from radio observations causing a reduction of 
the expected gamma-ray luminosity. However, even by assuming this unfavourable (and {\it ad hoc})
situation, we checked that the decrease of the gamma-ray luminosity with respect to our
calculations can be estimated within a factor $\leq$1.5-2 considering injection spectra of CRp
$\delta \geq 2$, thus it would not affect our main conclusions too much. 

At this point it is also useful to 
derive the maximum energy of CRp that is allowed
by the Fermi-LAT limits as a function of $\eta_B$.
In Figure 4 we show three relevant examples that are obtained 
assuming $\eta_B=0$, 0.3 and 0.5 and the
corresponding values of the minimum $B_0$ for which the 
gamma-ray flux equals Fermi-LAT limits (i.e., $B_0=6$, 13 and 21$\mu$G
respectively; Fig.~3, left).
We find that the ratio of the energy density of CRp and thermal ICM
increases with distance, this is
in line with independent other findings that attempt to match the
observed radio profile of the Coma cluster
with pure secondary models \citep[e.g.,][]{Zandanel:2014aa}. In practice the
decline with radius 
of the thermal targets for CRp-p collisions combined with the very broad
radial profile of the synchrotron
brightness requires a substantial amount of CRp in the external regions.
The ratio $\epsilon_{\mathrm{CRp}}/\epsilon_{\mathrm{ICM}}$ increases by one order of
magnitude from about 1\% in the core to 
about 10\% at 3 core radii, where most of the thermal energy budget is
contained. 
Despite the spatial profile of the magnetic field in the three models
is very different ($\eta_B$), we 
note that $\epsilon_{\mathrm{CRp}}/\epsilon_{\mathrm{ICM}}$ does not differ very much.
This is not surprising, because the magnetic field in these models is
strong, $B^2 \gg B_{cmb}^2$ (Fig. 3), implying
that a change of the magnetic field with distance does not lead to a
strong variation of the ratio between gamma and radio luminosity 
(Eq.~\ref{ratioradiogamma}).

\begin{figure}
\includegraphics[width=0.45\textwidth]{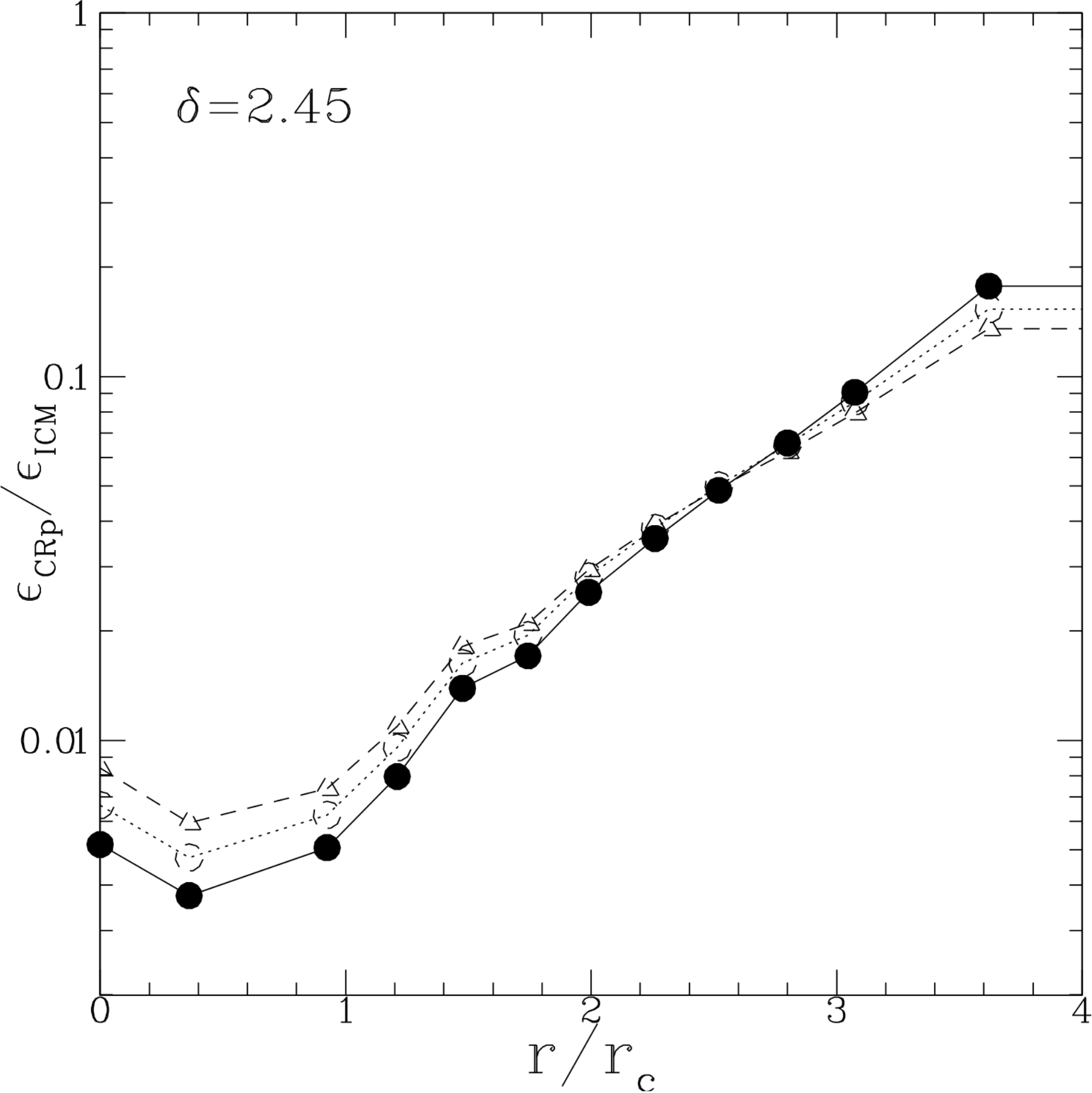}
\caption{Upper limits to the ratio of the CRp to the thermal ICM energy 
densities as a function of radius. The three models assumes $\eta_B=0$
(dashed), 0.3 (dotted) and 0.5 (solid) and the corresponding values 
of the lower-limit to the central magnetic fields $B_0$ from Figure 3 (left).}
\label{fig.5}
\end{figure}

\subsection{The general case with reacceleration}

In this section we consider the general case where 
turbulence is present in the ICM and reaccelerates CRp and 
secondary particles. We assume that the contribution of primary 
electrons is negligible (this is the case $f=1$ in Sect. 2) leading to 
optimistic expectations for the level of the gamma-ray emission.
Turbulent acceleration time is assumed constant with
radius, in practice this postulates that Mach numbers and turbulent 
injection scales (or their combination) are constant throughout the
cluster (see Sect.~6).

\begin{figure*}
\includegraphics[width=0.425\textwidth]{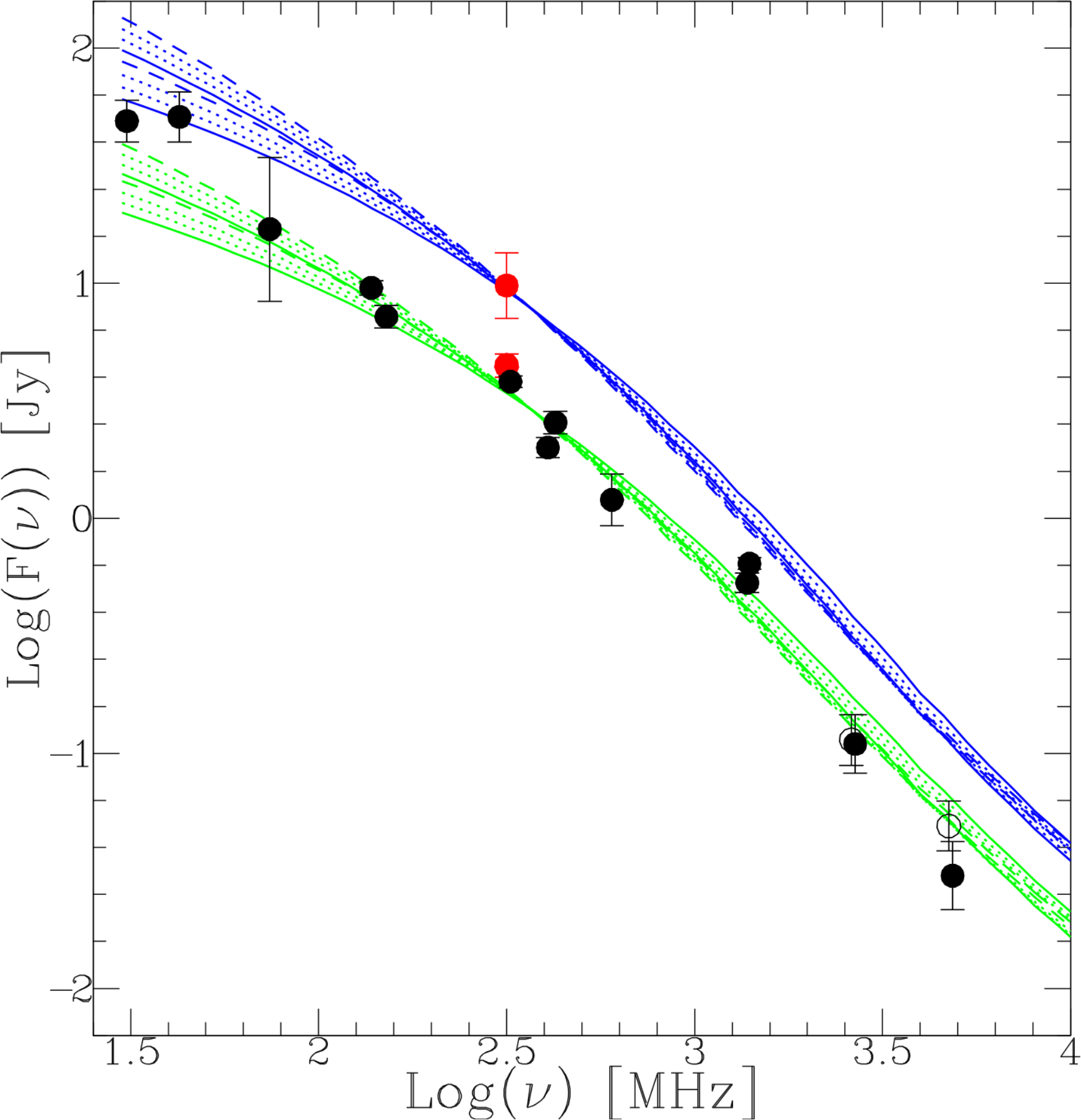}
\includegraphics[width=0.43\textwidth]{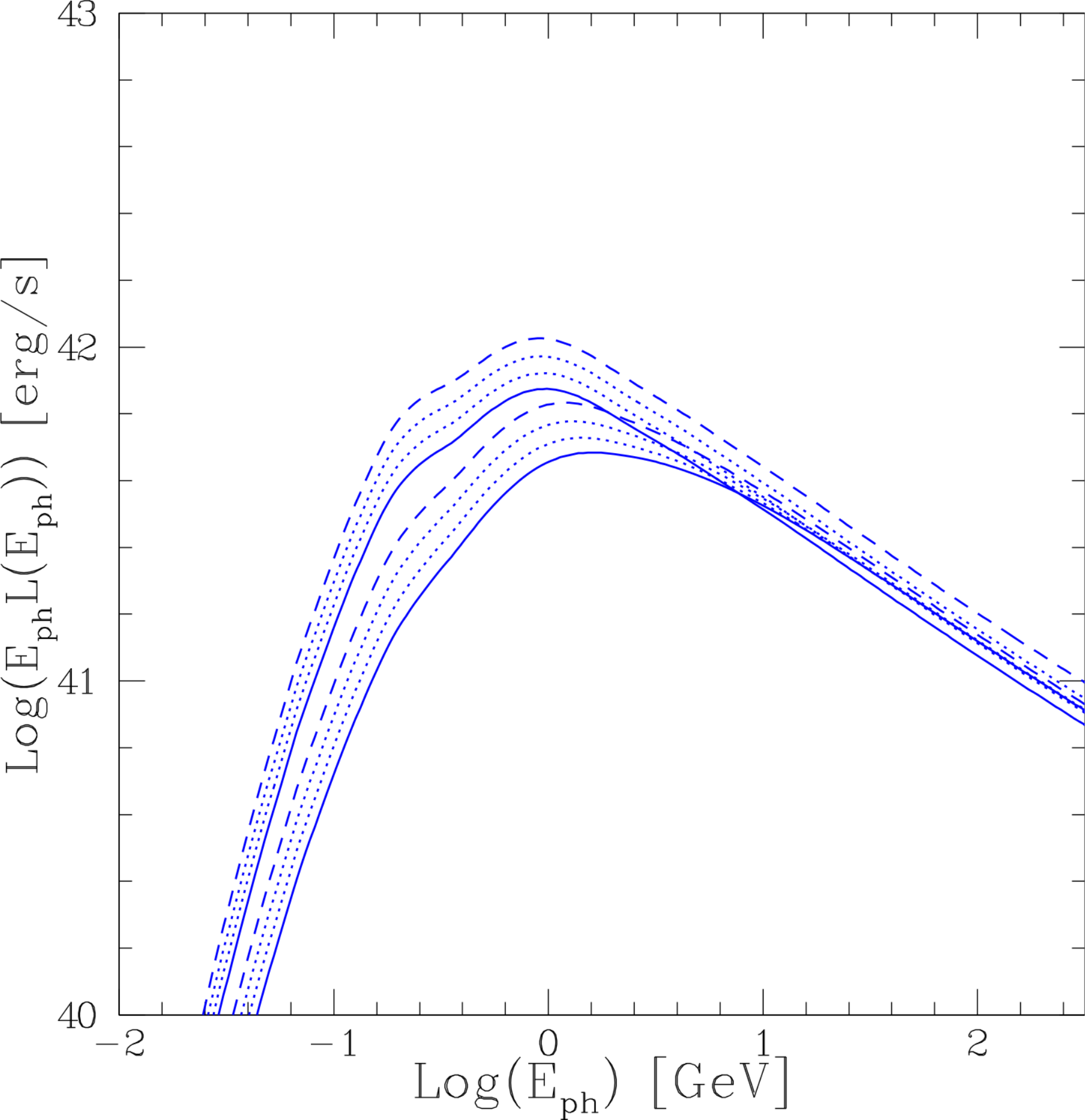}
\caption{Spectra emitted by a bundle of representative models.
(left) Radio spectrum from eigth reacceleration models obtained
assuming four reacceleration times ranging from $\tau_\mathrm{acc} = 230$ to 300
Myr and two reacceleration periods $\Delta T_\mathrm{acc}=$480 and $=$780 Myr. 
Calculations assume $z_i=0.2$.
The synchrotron flux is
calculated from models assuming two aperture radii $=$525 kpc 
(green, bottom) and 1.2 Mpc (blue, upper). 
Radio data (black points) are taken from
\citet{Pizzo:2010aa}; fluxes at high frequencies are corrected for the SZ
decrement considering a aperture radius $=$0.55 Mpc following
\citet{Brunetti:2013aa} (empty points at 2.7 and 4.8 GHz are the
original fluxes). The two radio fluxes obtained at 325 MHz using the
new WSRT data (red points) are extracted within apertures $=$0.6 (bottom) and
1.2 Mpc (upper) respectively.
(right) The gamma-ray spectrum obtained from the same models in the
left-panel calculated using an aperture radius $=1.2$ Mpc.}
\label{fig.6}
\end{figure*}

\noindent
First of all, we have to start from an initial spectrum of 
CRp in the cluster. As a simplification we assume that CRp in the ICM
are continuously injected between $z_i$ and $z_f$ with an injection spectrum 
$Q_p(p,t) \propto p^{-2.45}$. We assume that reacceleration 
starts at $z_f$ and that in the redshift range $z_i$ -- $z_f$ the CRp
are subject to Coulomb losses and CRp-p collisions (Sect. 3). 
In this time period we assume an average density of the thermal plasma 
$n_{\mathrm{ICM}} \sim 10^{-3}$cm s$^{-1}$.
In this Section we consider two reference cases $z_i=0.2$ and 0.5 and 
that reacceleration starts at a reference redshift $z_f=0.04$.

\noindent
Starting from the spectrum at $z=z_f$,  
we calculate the time evolution of CRp at several distances from the cluster 
center due to losses and reaccelerations and 
the production (and evolution) of secondary electrons 
and positrons.
In addition, turbulence reaccelerates also 
sub/trans-relativistic CRp to higher energies.
Thus the choice of different values of $z_i$ is aimed at
exploring the changes of the non-thermal emission that are induced by
assuming two ``extreme'' situations with 
initial CRp spectra that significantly differ at lower energies.
We note that
the exact value of the final redshift $z_f$ ($\pm$ 0.5 Gyr) 
does not significantly affect our results.

\noindent
At this point the free parameters of the models are:
\begin{itemize}
\item{the magnetic field configuration, $B_0$ and  $\eta_{B}$;}
\item{the acceleration efficiency $\tau_\mathrm{acc}^{-1}$ (which
depends on the combination of turbulent Mach number and injection
scale); and}
\item{the duration of the reacceleration phase, $\Delta
T_\mathrm{acc}$.}
\end{itemize}
\noindent
We proceed by assuming a magnetic field model $(B_0, \eta_{B})$ and
explore a range of values for $\tau_\mathrm{acc}^{-1}$ and $\Delta T_\mathrm{acc}$.
As explained in Sect.~3.3, in the case of collisionless TTD the CRp do not
contribute very much to the turbulent damping and thus they can be
treated as test particles in our calculations.
As a consequence, for each model 
the normalization of the spectrum of CRp
(essentially their number density) at each distance from the cluster center is
adapted to match the observed brightness distribution at 350 MHz.
This procedure provides a bundle of models, anchored to the radio
properties of the halo at 350 MHz, for which 
we calculate the synchrotron spectrum integrated within different 
aperture radii and the corresponding gamma-ray emission (spatial distribution 
and spectrum).

\noindent
Examples of 
the synchrotron spectra of the Coma
halo calculated for a bundle of models 
and integrated in an aperture
radius of 525 kpc (bottom) and 1.2 Mpc (top) are reported in Figure 5
(left). We note that 500-550 kpc is a typical 
value for the aperture radius that is used in the literature 
to derive the flux of the radio halo \citep[see discussion in][]{Brunetti:2013aa}.
Models in Fig. 5 are calculated assuming a magnetic field 
$B_0=4.7 \mu$G and $\eta_B=0.5$, i.e., the best fit parameters from Faraday
RM, \citep{Bonafede:2010aa}. 
Models have the same luminosity 
at 350 MHz as they are all constructed to match the brightness 
profile of the halo at this frequency. Considering the range of
parameters explored in Fig. 5, 
the models fit the observed curvature of the spectrum well, only showing 
deviations towards lower frequencies $\nu \leq 100$ MHz.
We also note that all these models explain the spectral curvature of
the (high-frequency) data in Fig. 5, 
which have already been corrected for the SZ decrement following \citet{Brunetti:2013aa}. 
This curvature {\it per se} demonstrates the existence of a break in the spectrum 
of the emitting electrons and has been used to support reacceleration models for the origin 
of the halo \citep[e.g,][]{ Schlickeiser:1987aa, Brunetti:2001aa}.

\noindent
In Figure 5 (right) we show the gamma ray spectrum generated by the
same models of Fig. 5 (left) integrating within a radius of
$=1.2$ Mpc.
While synchrotron spectra are fairly similar across a large 
range of frequencies, the corresponding 
gamma ray spectra generated between 
100 MeV and a few GeV vary greatly.
This is because, contrary to the synchrotron spectrum that is sensitive 
to the non-linear combination of electron spectrum and magnetic fields, 
the gamma rays directly trace the spectral 
energy distribution of CRp.
In addition, we note that
the GeV photons generated by the decay of neutral pions and
the synchrotron photons that are emitted by secondary electrons 
at $\geq$100 MHz do not probe the same energy-region of the spectrum of CRp.
Furthermore, the spectrum of secondary electrons are modified 
in a different way with respect to that of CRp during reacceleration 
\citep[e.g., Fig. 3 in][]{Brunetti:2011aa} and thus the ratio between GeV and radio
emission is also sensitive to the adopted reacceleration parameters.

\noindent
Figure 5 shows the potential of gamma-ray observations: they allow 
to disentangle among different models that otherwise could not be separable 
from their synchrotron spectrum alone.
Thus, following Sect.~4.2 we derive Fermi-LAT limits for a large
number of models calculated using a 
reference range of parameters and compare these limits to the 
gamma-ray emission expected from the corresponding models.
The procedure allows us to accept/rule out models and consequently to
constrain the parameter space $\tau_\mathrm{acc}^{-1}$, $(B_0, \eta_B)$,
$\Delta T_\mathrm{acc}$. More specifically we span the following range
of parameters:
\begin{itemize}
\item{a narrow range of 
$\tau_\mathrm{acc} = 200-300$ Myr (that is the typical range used to explain
the spectrum of the Coma halo); }
\item{a reasonable range of $\Delta T_\mathrm{acc} 
= 350-1000$ Myr (see Sect.~6); and }
\item{two families of magnetic field configurations 
with $\eta_B=0.5$ and 0.3 and $B_0$ as free parameters. 
Since models are 
normalized to the same radio flux at 350 MHz (anchored to the
brightness profile), 
the corresponding gamma-ray emission is expected to increase for 
smaller magnetic fields.}
\end{itemize}

\noindent
In Figure 6 we show the ratio of the flux above 100 MeV that is
expected from models and the LAT limits derived for the same
models as a function of the magnetic field in the cluster.
Here we assume $\tau_\mathrm{acc} = 260$ Myr and $\Delta T_\mathrm{acc}= 720$ Myr
that are representative, mean values of the range of parameters.
The gamma-ray flux increases for smaller values of $B_0$.
All models with $F_{>100}/F_{lim} \geq 1$ are inconsistent with LAT
observations.
In this specific case, we 
derive corresponding limits on the magnetic fields $B_0 \geq 4 \mu$G and 
$\geq 2.5 \mu$G in the case $\eta_B =0.5$ and $=0.3$, respectively. 
These limits are consistent with
the magnetic field values derived from Faraday RM.
For comparison the limits derived in the case of pure
hadronic models are much larger,
$B_0 \geq 21 \mu$G and $\geq 13 \mu$G, respectively (Sect.~5.1). 
In Fig. 6 we also show the threshold that is expected from upper limits
in the case of 10 and 15 yrs of Fermi-LAT observations (see the Section~6.1 for 
details on how the expected limits were computed).
The limits derived using 10-15 yrs of Fermi-LAT are expected to 
start constraining 
magnetic field values that are larger than (i.e., potentially
inconsistent with) those from RM.

Before making a more general point on the constraints on the magnetic
field that are allowed by LAT limits in the case of reacceleration
models, we first investigate the dependencies on $\tau_\mathrm{acc}$ and the
acceleration period.
In Figure 7 we show the ratio of the flux expected above 100 MeV 
and the Fermi-LAT limits as a function of reacceleration efficiency.
Here we assume the central value $\Delta T_\mathrm{acc}= 720$ Myr (solid lines)
and the best fit
values from Faraday RM ($B_0 = 4.7 \mu$G and $\eta_B=0.5$).
The gamma-ray luminosity (and also $F_{>100}/F_{lim}$) increases with 
$\tau_\mathrm{acc}$, i.e., models with less efficient reacceleration produce
more gamma rays. This trend does not depend on the specific choice
of magnetic field configuration and is due to a combination of effects.
First, the spectrum of
CRp is less modified (i.e., less stretched in energy) in models with slower 
reacceleration implying a larger number of CRp in the energy range
1-10 GeV. Second, the spectrum of radio-emitting electrons is less boosted 
in the case of models with slower reacceleration implying that a larger 
number density of CRp is required to generate the observed flux at 350 MHz.
More specifically we find that the latter effect dominates within
about 2-3 
reacceleration times, after which the spectral shape of the emitting
electrons reaches quasi-stationary conditions and the evolution of the 
gamma-ray to radio ratio evolves mainly due to the 
modifications of the spectrum of CRp. 
In Fig. 7 we also show the threshold resulting from upper limits
assuming 10-15 yrs of Fermi-LAT data. Assuming 
magnetic fields from the best fit of Faraday RM analysis, the expected 
limits considering an observation of 10-15 years of continued LAT exposure are poised to put
tension on the majority of the range of acceleration times and acceleration 
periods explored in this paper (see the Section~6.1 for 
details on how the expected limits were computed).
\noindent
Coming back to the minimum value of the magnetic fields that can be assumed 
without exceeding observed gamma-ray flux limits, 
we note that the increase of the gamma-ray luminosity with $\tau_\mathrm{acc}$ 
implies that the minimum value of the 
central magnetic field $B_{0}$ increases with $\tau_\mathrm{acc}$.

\begin{figure}
\includegraphics[width=0.45\textwidth]{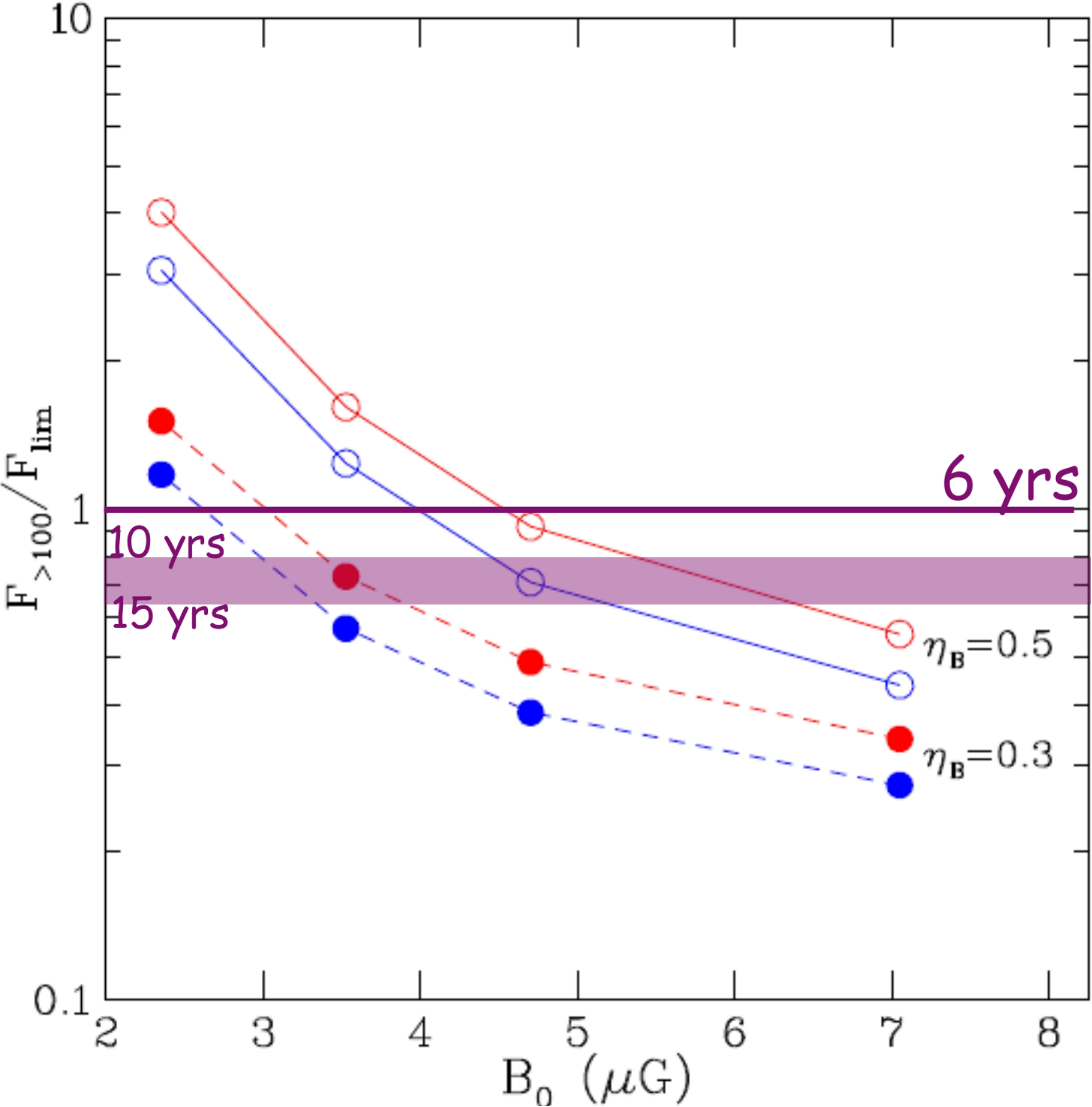}
\caption{The ratio of the flux predicted by models for $\geq$100 MeV and the
Fermi-LAT upper limits (obtained after 6 yrs of data in the same
energy band) is shown as a function
of the central magnetic field assumed in the models.
Calculations assume $\tau_\mathrm{acc}=260$ Myr, $\Delta T_\mathrm{acc}=720$ Myr,
$z_i=0.2$ (red) and $z_i=0.5$ (blue), and $\eta_B=0.3$ (dashed) and
$\eta_B=0.5$ (solid).
The horizontal pink region marks $F_{>100}/F_{lim}=1$ expected after 10
and 15 yrs of Fermi-LAT data.}
\label{fig.7}
\end{figure}

The other relevant parameter is $\Delta T_\mathrm{acc}$.
Once the synchrotron spectra are normalized to the observed luminosity
at 350 MHz, shorther reacceleration periods 
produce more gamma rays simply because \lumi
increases with time.
This is shown quantitatively in Fig.7 where we compare the case $\Delta T_\mathrm{acc}=$720
Myr with a shorther reacceleration period, $\Delta T_\mathrm{acc}=480$ Myr.
\noindent
The gamma-ray luminosity scales in the opposite way
with $\tau_\mathrm{acc}$ and $\Delta T_\mathrm{acc}$.
Faster reacceleration in shorther times may produce spectra that are 
comparable to the case of slower reacceleration in longer times
inducing a corresponding degeneracy also in the expected gamma-ray
flux. This is also clear from Fig.7: slower reacceleration with
$\Delta T_\mathrm{acc}=$720 Myr generates about the same gamma-ray flux in 
the case of faster reacceleration with $\Delta T_\mathrm{acc}=$480 Myr.

\begin{figure}
\includegraphics[width=0.45\textwidth]{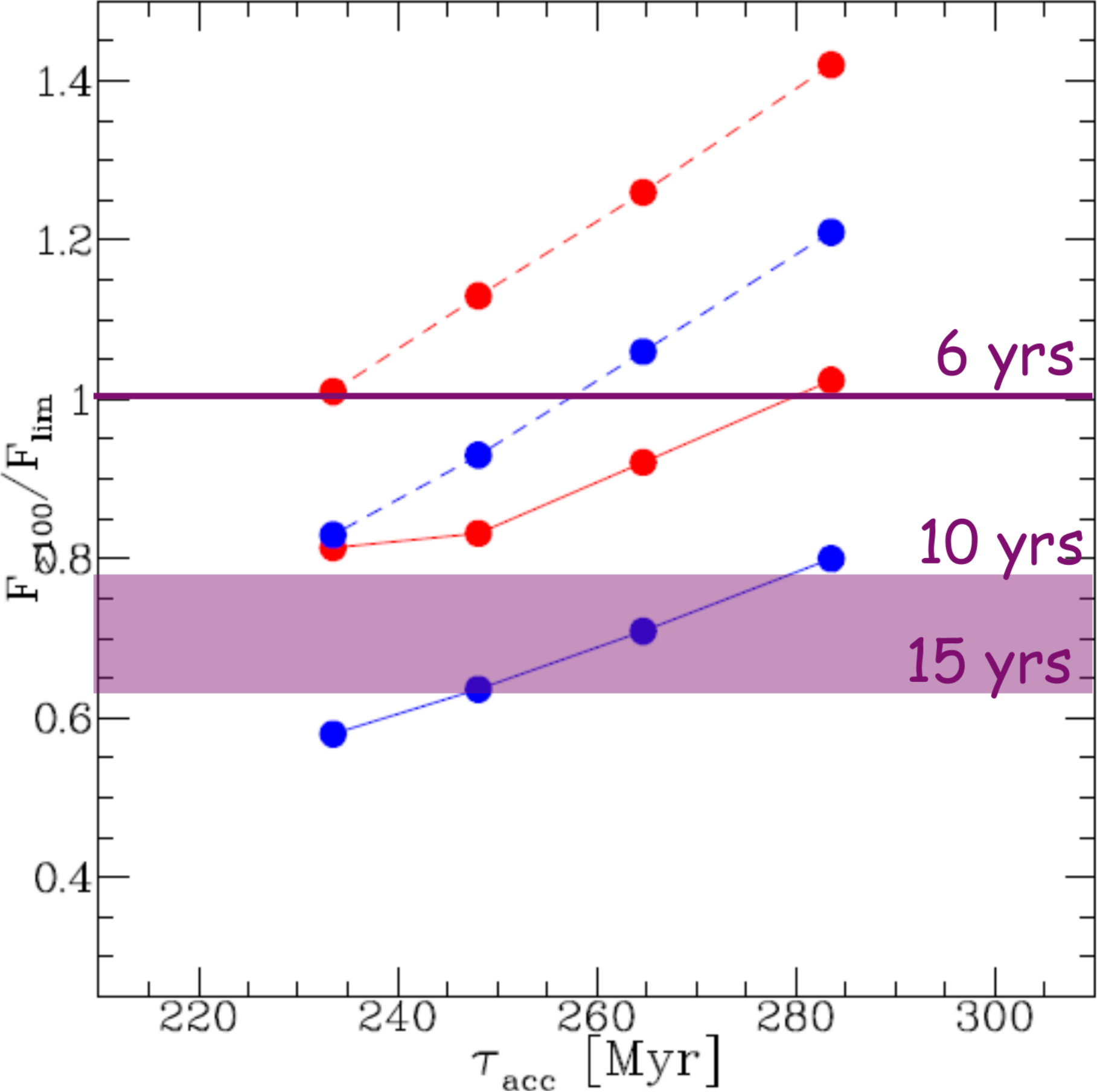}
\caption{The ratio of the flux predicted by models for $\geq$100 MeV
and the Fermi-LAT upper limits (obtained after 6 yrs of data in the same
energy band) is shown as a function
of the acceleration time.
Calculations assume $\eta_B=0.5$, and $B_0=4.7 \mu$G, 
$\Delta T_\mathrm{acc}=720$ Myr (solid) and 480 Myr (dashed),
and $z_i=0.2$ (red) and $z_i=0.5$ (blue).
The horizontal pink region marks $F_{>100}/F_{lim}=1$ expected after 10
and 15 yrs of Fermi-LAT data.}
\label{fig.8}
\end{figure}

The combination of radio and gamma-ray limits allows a combined
constraint on $\tau_\mathrm{acc}$ and $\Delta T_\mathrm{acc}$.
In Figure 8 we show the minimum values of the reacceleration period
that is constrained by gamma-ray limits as a function of reacceleration
time. For consistency with previous Figures, this is also
derived assuming the best fit values from Faraday RM.
Note that smaller values of $\Delta T_\mathrm{acc}$ (for a given $\tau_\mathrm{acc}$)
can be allowed assuming larger 
magnetic fields in the cluster. 

Figs.~6-8 show that lower limits 
on the magnetic fields in the cluster increase with increasing 
$\tau_\mathrm{acc}$ 
and with decreasing $\Delta T_\mathrm{acc}$ and that these limits are 
not very far from the field values derived from Faraday RM.  
A relevant example 
is shown in Figure 9 where we report the minimum value of the
magnetic field as a function of the acceleration time and for two
values of the acceleration period. Calculations are obtained assuming
$\eta_B=0.5$ and are compared with the value of $B_0$ that is derived
from RM for the same $\eta_B$. 
Fig.~9 clearly shows that models with $\Delta T_\mathrm{acc} \leq 400-450$ Myr
are in tension with current constraints from Faraday RM because a
magnetic field that is too large is required in these models in order to
avoid exceeding the gamma-ray limits.
On the other hand, the minimum values of $B_0$ that are constrained for 
models with longer reacceleration periods become gradually consistent 
with Faraday RM. Interestingly, we conclude that the gamma-ray flux 
that is expected from reacceleration models assuming the magnetic field 
configuration that is derived from RM is similar (within less than 
a factor 2) to current Fermi-LAT limits, a fact that is particularly
important for future observations as we outline in the Discussion
section.

As a final point in Figure 10 we show the 
ratio $\epsilon_{\mathrm{CRp}}/\epsilon_{\mathrm{ICM}}$ as a
function of radius for reacceleration models that assume 
$\tau_\mathrm{acc}=260$ Myr and $\Delta T_\mathrm{acc}=720$ Myr, and assuming different 
configurations of the magnetic field (see caption).
We find that the ratio of the energy density of CRp and thermal ICM
increases with distance. The increase of
$\epsilon_{\mathrm{CRp}}$ with radius is faster than in the case of pure
hadronic models (Fig. 4) and also the differences between the case $\eta_B=0.3$ 
and 0.5 are more pronounced. This can be explained by the 
magnetic fields in reacceleration models which are much smaller than those
allowed in the case of pure secondaries implying that differences
in the radial decline of $B$ with radius induce significant changes in
the synchrotron emissivity.
We note that the models in Figure 10, that assume
$B_0=2.35~\mu$G and $\eta_B=0.3$ and
$B_0=4.7~\mu$G and $\eta_B=0.5$, mark the situations where the 
resulting predicted gamma-ray fluxes would be in tension with the observed limits, 
implying that the corresponding values of $\epsilon_{\mathrm{CRp}}/\epsilon_{\mathrm{ICM}}$ are upper limits. 
In fact these limits 
constrain the ratio $\epsilon_{\mathrm{CRp}}/\epsilon_{\mathrm{ICM}}$ at distance 
$r \sim 3 r_c$ to about 7-8$\%$ in both cases.
The fact that the limit to the amount of CRp energy
is slightly (factor 2) larger in the case of pure secondary models
(Fig.~4) is due to the changes in the spectral shape of CRp (and
gamma rays) derived from reacceleration (Fig.~5 right),
and due to the differences between the spatial distribution of CRp in the
two models. 

\begin{figure}
\includegraphics[width=0.44\textwidth]{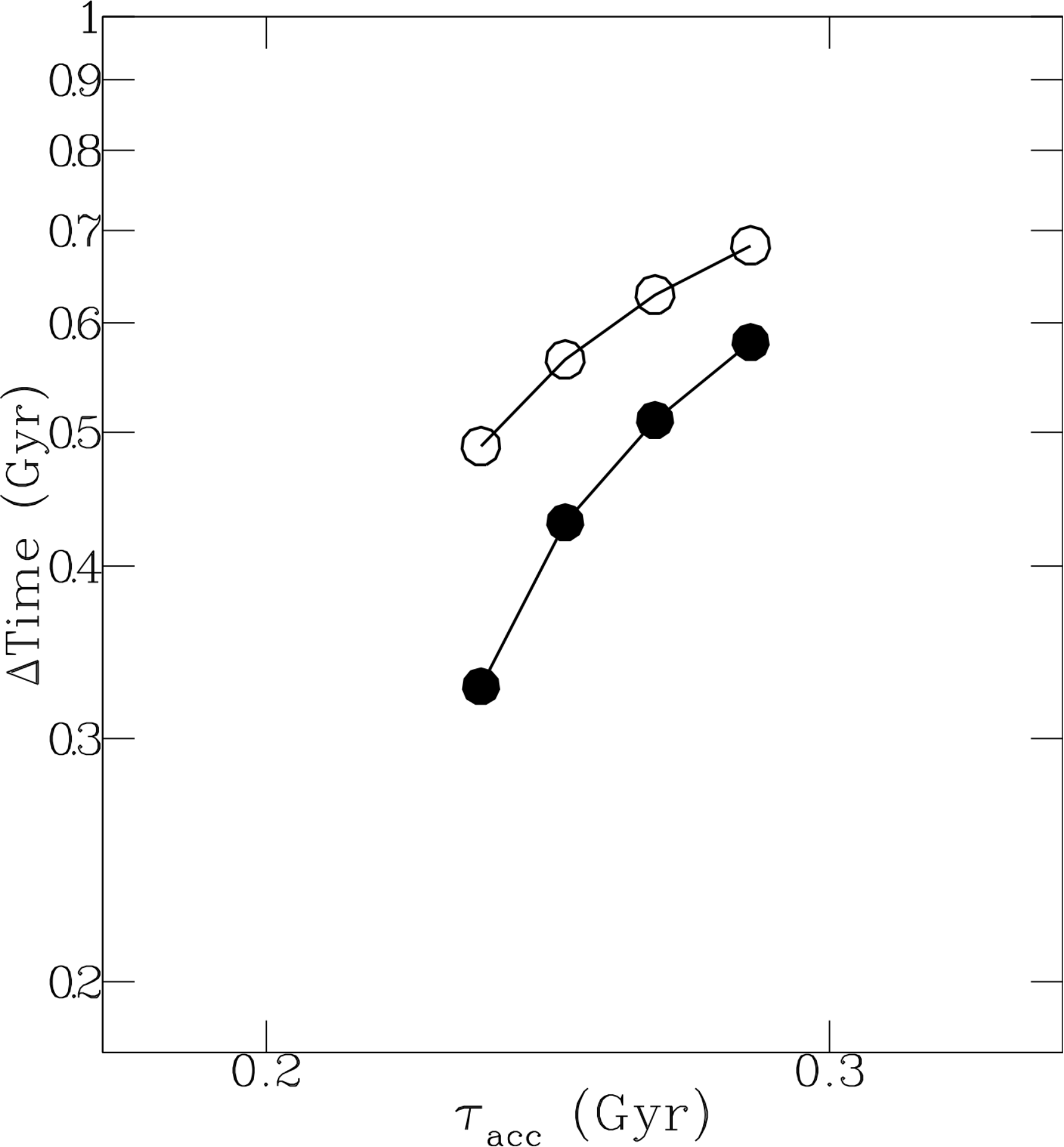}
\caption{The lower limits to $\Delta T_\mathrm{acc}$ as a function of
$\tau_\mathrm{acc}$ are obtained comparing the flux expected by models
with the Fermi-LAT limits. Calculations are obtained for
$\eta_B=0.5$ and $B_0=4.7 \mu$G and assuming $z_i=0.2$ (upper) and
$=$0.5 (lower).}
\label{fig.9}
\end{figure}

\begin{figure}
\includegraphics[width=0.45\textwidth]{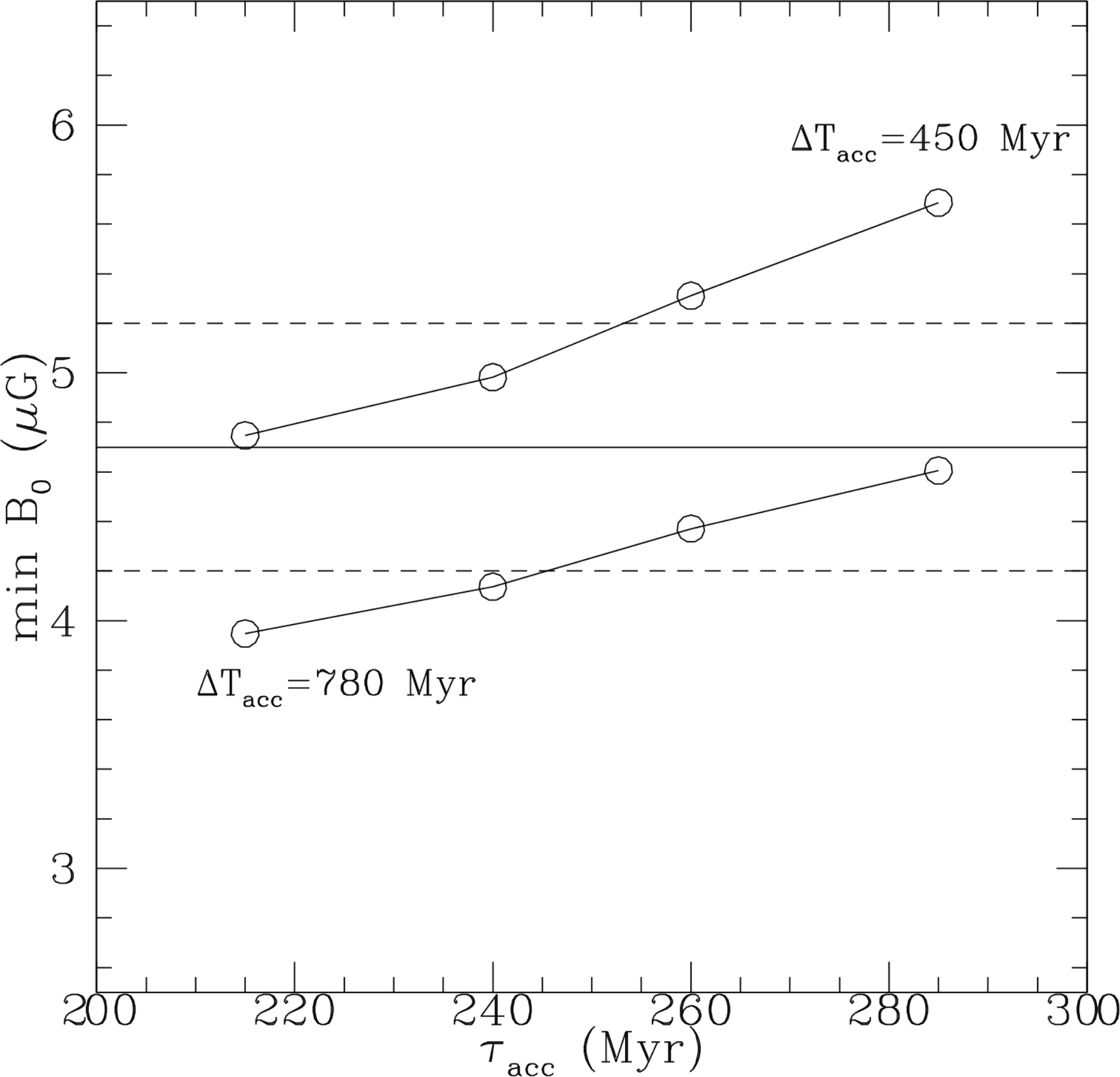}
\caption{Lower limits to the central magnetic field, in the case
$\eta_B=0.5$, as a function of reacceleration time.
Calculations assume $z_i=0.2$, and $\Delta T_\mathrm{acc}=450$ Myr (upper) and $=$780 Myr
(lower curve). Limits are compared to the range of values $1 \sigma$
(dashed lines) constrained by Faraday RM \citep{Bonafede:2010aa}.}
\label{fig.10}
\end{figure}

\begin{figure}
\includegraphics[width=0.45\textwidth]{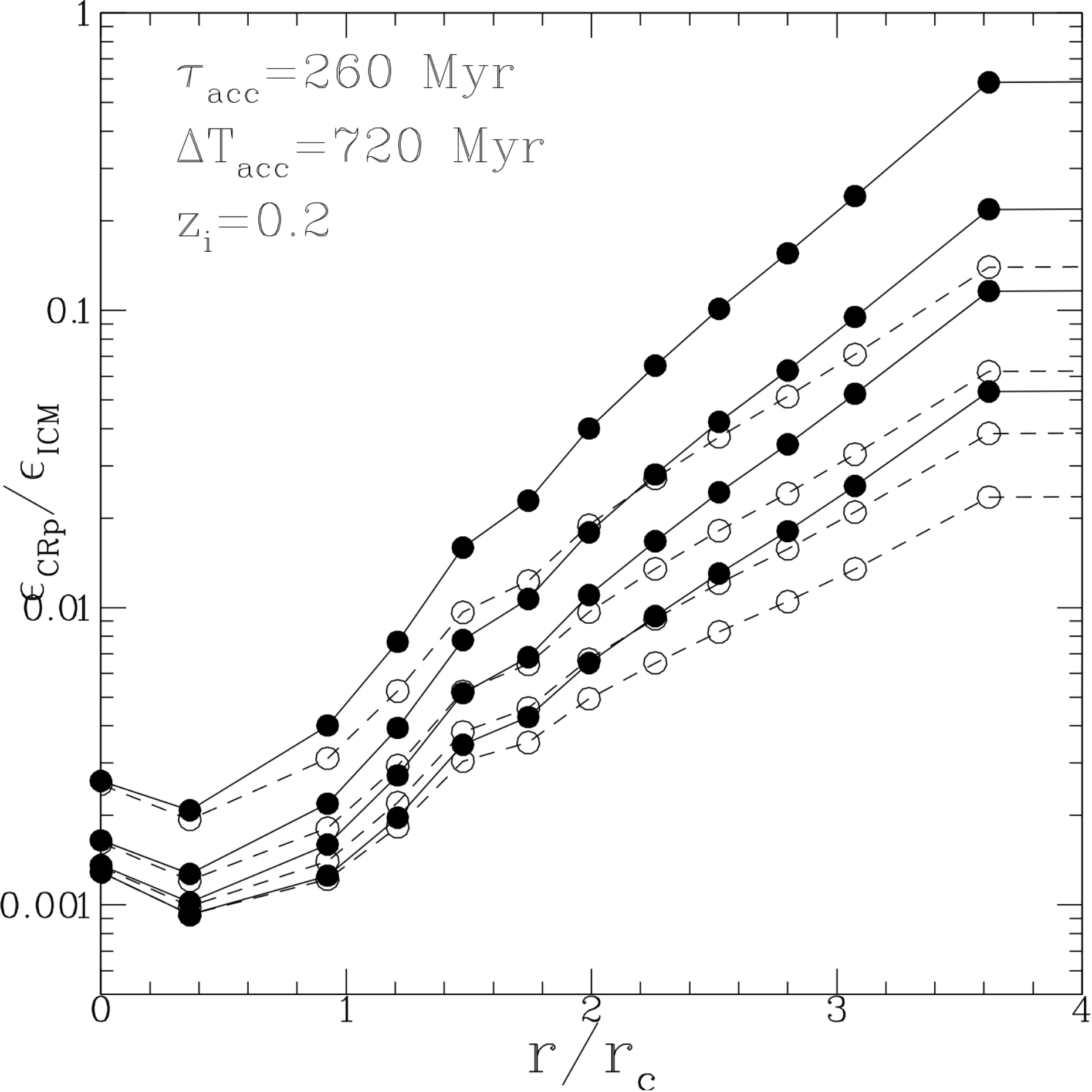}
\caption{Upper limits to the ratio of the CRp to the thermal ICM energy
densities as a function of radius. Models assume $\tau_\mathrm{acc}=260$ Myr,
$\Delta T_\mathrm{acc}=720$ Myr and $z_i=0.2$.
Results are shown for the cases $\eta_B=0.5$ (solid lines) and
$=$0.3 (dashed), and considering four values of the central
magnetic fied: $B_0= 0.5\times$, $0.75\times$, $1.0\times$ and
$1.5\times 4.7 \mu$G (from bottom to upper lines).}
\label{fig.11}
\end{figure}

\section{Discussion}

In this section we expand the discussion of our results including limitations of the current approach.

\subsection{A detection in next years?}

Our study demonstrates that turbulent
reacceleration of secondary particles can explain the observed radio
halo without exceeding the current gamma-ray limits. 
However, we have shown that
assuming a magnetic field in the cluster that is 
compatible with Faraday RM, the level of the resulting gamma-ray flux 
is similar to what is currently constrained by observations of Fermi-LAT.
This conclusion opens up the interesting perspective that
the role of CRp for the observed cluster-scale radio emission may be
efficiently tested with future observations.

\begin{figure}
\includegraphics[width=0.45\textwidth]{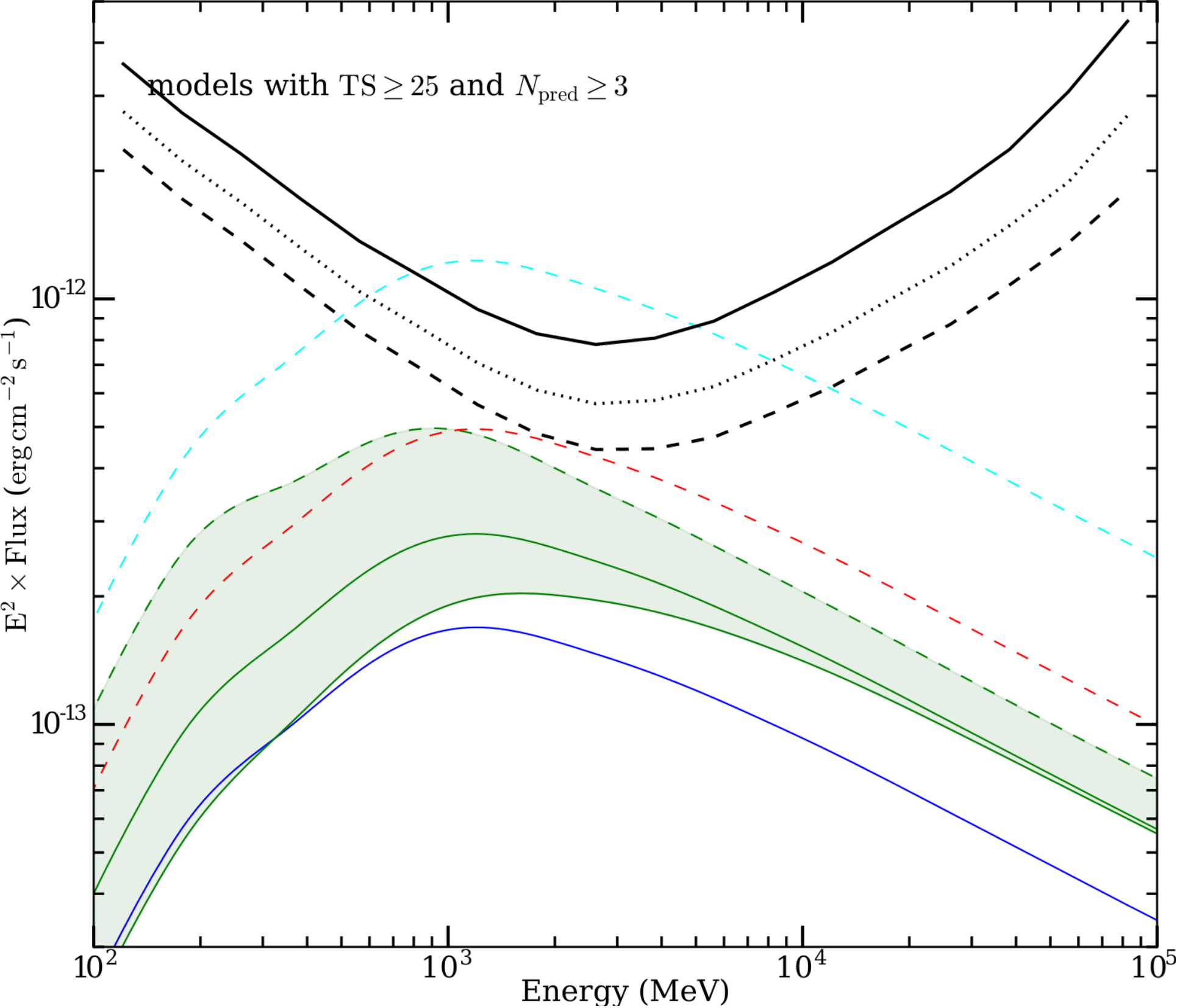}
\caption{Predicted spectra (corrected for extension, see text)
are compared with the sensitivity curves 
(concave black lines) of the Fermi-LAT after 6 (solid), 10 (dotted)
and 15 yrs (dashed) of observations.
Models assume $z_i=0.2$ and $\eta_B=0.5$.
Two families of calculations are shown: 
(1) four models assuming $\tau_\mathrm{acc}=260$ Myr, $\Delta T_\mathrm{acc}=720$
Myr and the four central magnetic field values $B_0= (0.5, 0.75,
1.0, 1.5)\times 4.7 \mu$G (cyan dashed, red dashed, green solid (upper one)
and blue solid, respectively);
(2) two models assuming a central magnetic field $B_0=4.7 \mu$G and
the two combination of parameters $\tau_\mathrm{acc}=290$ Myr and $\Delta
T_\mathrm{acc}=470$ Myr (red dashed),
and $\tau_\mathrm{acc}=210$ Myr and $\Delta T_\mathrm{acc}=760$ Myr (green solid,
lower one).
The green region encompasses the range of spectra that are produced
assuming a magnetic field configuration from the best fit of Faraday RM,
$B_0=4.7 \mu$G and $\eta_B=0.5$.}
\label{fig.12}
\end{figure}

\noindent
Turning the latter argument around, this also open to the possibility 
of a detection of the Coma cluster in
the next years of observations, provided that CRp play an
important role for the origin of the observed radio halo.
To check this point we consider the sensitivity with 10 and 
15 years of integrated sky exposure.
In Fig. 11 we show the median sensitivity as provided 
by the so-called ``Asimov'' data set \citep{Cowan:2011aa}
assuming that the background is determined entirely by both the Galactic 
diffuse emission model along with the isotropic extragalactic model, both
provided by the LAT collaboration and made available through the 
Fermi Science Support Center (FSSC).
\footnote{\url{http://fermi.gsfc.nasa.gov/ssc/data/access/lat/BackgroundModels.html[4]}, 
{\tt{gll\_iem\_v06.fits}} and {\tt{iso\_P8R2\_SOURCE\_V6\_v06.txt}}, respectively} 
For its calculation we assume that the test
statistic $TS$, given by the likelihood ratio between the null-hypothesis (the gamma-ray emission 
observed in the cluster field consists of Galactic and isotropic extragalacticemission) 
$\mathcal{L}_{0}$ and the alternative hypothesis
$\mathcal{L}_{1}$ (null-hypothesis in addition to the cluster emitting gamma rays according to the tested model) is larger than 25, which can be related to Gaussian
significance $n$ as $n=\sqrt{TS}$, as well as requiring at least 3
photons to be attributed to the signal.\footnote{We loosely refer to the case of 
$TS\gtrsim25$ as \emph{detection} and assume that the underlying null-distribution 
is $\chi^{2}$-distributed.} 
This sensitivity calculation is implemented in {\it FermiPy} which is used with the corresponding set 
of IRFs and evaluated at the sky position of the
Coma cluster.\footnote{FermiPy (\url{http://fermipy.readthedocs.io/en/latest/}) provides a set of convenience 
wrappers and a unified python interface to the Fermi-LAT Science tools provided by the FSSC.} 
Note that the sensitivity curve is calculated assuming the emission to be point-like. 
In order to provide a reasonable
comparison, we correct each of the model spectra for the extension
by calculating a correction factor given as the ratio of the
the uppr limit assuming the emission to be point-like
and the flux upper limit under the model assumption. 

\noindent
In Fig. 11 we compare the sensitivity curves to the (corrected)
expected gamma-ray
spectra obtained assuming $\tau_\mathrm{acc} = 260$ Myr, $\Delta
T_\mathrm{acc}=720$ Myr, $\eta_B=0.5$ and $B_0=(0.5, 0.75, 1.0, 1.5)\times
4.7 \mu$G (from top to bottom); this is the same model
configuration in Figure 6 (upper red solid line).
High TS-values, $TS > 25$, are predicted only for models that
are already in tension with current upper limits (reported with dashed
lines for sake of clarity), however there is a
chance to obtain a signal in excess of $TS=9 \simeq 3\sigma$ also for
models that are still compatible with current limits. As a reference
case, in Fig. 11 we show the spectral region (green) that is spanned 
by models with the magnetic field from the best fit from RM 
analysis ($B_0=4.7 \mu$G and $\eta_B=0.5$) and assuming the full range 
of model parameters adopted in our paper.

\subsection{Contribution from primary electrons in reacceleration models}

In all the calculations carried out in this paper we have assumed $f=1$, 
i.e., the case of a negligible contribution to the radio emission
from reaccelerated primary electrons.
Thus our calculations are suitable to describe situations where
the contribution of primary electrons is subdominant, similar to the
case observed in our Galaxy.
As discussed in Sect. 2, the gamma-ray flux essentially scales
with $f^{-1}$ and consequently 
the case $f=1$ leads to the maximum gamma-ray
emission that is predicted from reacceleration models for a given
reacceleration efficiency and configuration of the magnetic field.

\noindent
The ratio between primary and secondary particles in the ICM is
unknown. Theoretical arguments suggest that CRp are the most 
important non-thermal component in galaxy clusters \citep[][for
review]{Blasi:2007aa,Brunetti:2014aa}, implying that secondaries
might indeed play a role. However this appears to
be very sensitive to the acceleration efficiency
at cluster shocks, which is poorly known
\citep[e.g.,][]{KangRyu:2011aa,KangRyu:2013aa,Caprioli:2014aa}.
As a consequence 
it may very well be that $f\gg 1$ and that 
the gamma-ray flux from the Coma cluster is
much smaller than that predicted by our calculations.
In fact, originally the reacceleration models have been proposed 
assuming turbulent reacceleration of primary seed-electrons that may
naturally accumulate at energies of about 100 MeV in the ICM as a result
of the activity of shocks, AGN and galaxies 
\citep[e.g.,][]{Brunetti:2001aa,Petrosian2001}.
Most cosmological simulations point toward a dominant CRp 
contribution in clusters centre and an increasingly important role
of primary electrons in clusters outskirts 
due to the complex network of shocks typically present there 
\citep[e.g.,][]{2008MNRAS.385.1211P,Skillman:2008aa,Vazza:2009aa,
2010MNRAS.409..449P} leading to an increasing ratio 
$f$ with radius. This could possibly also alleviate the issue of the 
distribution of CRp needed to match giant radio halos, 
like Coma, that is required to significantly increase with radii
\citep[e.g.,][, this work]{Brunetti:2012aa,Zandanel:2014ab}.
In this respect we also mention that several possible solutions to match
the broad radial profiles of the Coma halo include CRp streaming,
an enhanced efficiency for electrons
acceleration at shocks, or an increasing turbulent-to-thermal 
energy ratio with radii \citep{Pinzke:2017aa}. 

\noindent
The detection of gamma-ray emission from the Coma cluster in the future will constrain $f$ of the order of
unity (assuming a magnetic field configuration in line with RM),
however a non-detection at the sensitivity level that
Fermi-LAT will reach in 10-15 years would not necessarily imply that
reaccelerated secondary particles are completely subdominant.

\subsection{Assumptions in acceleration models and parameter space}

Our conclusions are based on the exploration of a
limited range of model parameters and on a number of assumptions.
While the former constitutes no significant limitation, evaluating in detail what changes 
are induced by relaxing some of our assumptions requires a follow-up study and is beyond 
the scope of this work. We comment on both points in the following.

The first parameter in our model is the acceleration time $\tau_\mathrm{acc}$ 
that we assume in the range 200-300 Myr.
In the situation where reacceleration time is significantly shorter than 
the lifetime of the halo (or reacceleration period) the acceleration 
efficiency sets the maximum (i.e., steepening)
synchrotron frequency of the spectral component 
that is powered by turbulent reacceleration.
For constant reacceleration in the volume, the maximum frequency is typically 
generated in the regions of the cluster where 
$B \sim B_\mathrm{IC}/\sqrt{3}$. At the redshift of the Coma cluster
(where $B_\mathrm{IC}/\sqrt{3} \simeq 2 \mu$G) this steepening
frequency is \citep[from Eq.~14 in][]{Brunetti:2016aa} 
$\nu_s/{\rm GHz} \sim ( \tau_\mathrm{acc}/400 {\rm Myr} )^{-2}$.
This requires $\tau_\mathrm{acc} \sim 300$ Myr to explain the observed spectral
steepening at $\sim 1-2$ GHz. Only if the magnetic field strength in the
cluster core is sub-$\mu$G the required acceleration time should be 
$< 200$ Myr. In this case, however, the gamma-ray luminosity would be
strongly in excess of the Fermi-LAT limits, and such weak fields would 
also be much smaller than those derived from RM.
We thus conclude that our calculations essentially
span the range of values of $\tau_\mathrm{acc}$ that is relevant to explain the
radio spectral properties of the halo.

The other parameter is the acceleration period, $\Delta T_\mathrm{acc}$, that 
we assume in the range 350-1000 Myrs. 
Specifically our conclusions depend on the ratio between the gamma-ray 
and radio luminosities which essentially depends on the reacceleration 
rate (and losses) that {\it is experienced by secondary particles in the 
last} $\Delta T_\mathrm{acc}$. $\Delta T_\mathrm{acc}$ is presumably
smaller than the age of the radio halo which in fact can be
longer than 1 Gyr \citep[see, e.g.,][]{Miniati:2015ab,Cassano:2016aa}.
Indeed it is very unlikely that
particles in the ICM are continuously reaccelerated at a constant rate
for much more than a turbulent cascading time, that is a few hundred
Myr, and this motivates the range of $\Delta T_\mathrm{acc}$ that we assumed in
our calculations.
Even if turbulence in the ICM is long living, on longer times the dynamics
of ICM would transport and mix CRp and their secondaries on large scales 
implying that CRs sample different physical conditions and different
reacceleration efficiencies. 

In our calculations we have assumed that the acceleration efficiency 
is constant in the cluster volume.
In the case of TTD and under the assumption that turbulent interaction with both thermal ICM and 
CRs is fully collisionless (Sect. 3) the acceleration time
is \citep{Brunetti:2016ab}:
\begin{equation}
\tau_\mathrm{acc} = {{p^2}\over{4 D_{pp}}}
\sim 125
({{M_o}\over{1/2}})^{-4} \left(
{{L_o/300 \, {\rm kpc}}\over{c_s/1500\, {\rm km \,s^{-1}}}}
\right) \,\,({\rm Myr})
\label{timeaccC}
\end{equation}
\noindent
where we assumed a Kraichnan spectrum for MHD turbulence.
Consequently a constant acceleration time implies (neglecting the
dependence on temperature) that $L_0/M_o^4$ is constant.
However numerical simulations suggest that turbulence in the ICM is
stronger outside cluster cores and that the turbulent pressure is more
relevant in the external regions \citep[e.g.,][]{Vazza:2017aa}.
If the turbulent Mach number increases radially, this would result
in higher acceleration rates in the external regions \citep[see
also][]{Pinzke:2017aa}.
In this case
matching the observed synchrotron profile would require an energy
budget of the CRp in the external regions that is slightly smaller 
with the consequence that a lower gamma-ray luminosity is predicted.

\noindent
From Eq.~\ref{timeaccC} it emerges that compressive turbulence in the
Coma cluster should have a Mach number $\sim 0.4 ( L_0/300 )^{1 \over 4}$ 
to guarantee the acceleration rates $\tau_\mathrm{acc} \sim 200-300$ Myr that have
been assumed in our calculations. 
This would result in about $15 \%$ (or $5 \%$) of 
pressure support from compressive motions generated 
on 300 kpc (or 30 kpc) scales. 
Direct measurements of turbulent velocities are currently available only
for the cool-core of the Perseus cluster, where radial - line-of-sight -
turbulent velocities $\sim 160$ km$/$s on $L_0 \sim 30$ kpc scales have 
been derived from the analysis of the X-ray lines by the {\it Hitomi} 
collaboration \citep{Hitomi:2016aa}.
This would imply a turbulent pressure support (limited to the  
eddies on $L_0 \leq 30$ kpc scale) of about 4 \% of the thermal pressure.
Merging clusters - such as Coma - should be more turbulent 
\citep[e.g.,][]{Paul:2011aa,Vazza:2011aa,Nagai:2013aa,
Miniati:2014aa,Miniati:2015aa,Iapichino:2017aa}.
For example, coming back to Coma, a recent analysis that combines numerical
simulations and SZ fluctuations observed by Planck in the Coma cluster 
concluded for a dynamically important pressure support from
turbulent fluctuations, presumably in the form of adiabatic modes on
large scales \citep{Khatri:2016aa}.

Finally we note that in calculations of reacceleration models we assumed an injection
spectrum of CRp $\delta=2.45$, that is the reference spectrum that we have adopted 
in the case of pure hadronic models. 
On one hand this allows us to promptly establish the effect on the
SED that is driven by adding turbulence in hadronic models, on the
other hand this is a limitation of the current paper. 
Indeed the spectrum of secondary electrons in reacceleration models is not strightforwardly 
determined by the spectrum of CRp but rather by the combined effects
of injection and spectral modifications that are induced by reacceleration and energy losses. 
Consequently a different injection spectrum of CRp can still fit the 
observed synchrotron spectrum of the Coma halo provided that the
turbulent reacceleration rate is tuned accordingly. 
A full exploration of parameters in reacceleration models, including
different CRp injection spectra, is beyond the focus of the paper and
is a task for a forthcoming paper (Zimmer et al., in prep.). 
In principle steeper CRp spectra have a tendency to generate
significantly more gamma rays in the Fermi-LAT energy 
range. However our calculations are also constrained by the observed
spectrum of the radio halo in which case two combined effects compensate this tendency.
First, steeper CRp spectra also generate more secondary electrons with
$\sim$MeV--GeV kinetic energies that can be reaccelerated by turbulence leading to an 
increasing ratio between radio and gamma-ray luminosities which in turn implies a lower 
gamma-ray luminosity.
Second, in the case of steeper CRp spectra slightly shorter $\tau_{acc}$ (or longer 
$\Delta T_{acc}$) are required to fit the radio spectrum of Coma and these configurations in 
reacceleration models have the tendency to produce less gamma-rays (Sect. 5.2).
As a net result the gamma-ray luminosities in the Fermi-LAT band that
are generated by reacceleration models using $\delta$ in the range 2.2--2.6 are 
expected to differ within less than a factor 2.

\subsection{Beyond the case of collisionless TTD reacceleration}

In our calculations we assumed a fully collisionless TTD interaction between
compressive turbulence and both thermal ICM and CRs.
This leads to a conservative estimate of the
turbulent acceleration rate simply because in this way most of the
turbulent energy is channelled into the thermal ICM (heating of the plasma).
On the other hand this situation is very convenient because it allows
us to treat CRs as {\it passive} tracers in the turbulent field as
they do not contribute too much to turbulent damping (see Sect.~3).
On the other hand it is also possible that only CRs interact in a
collisionless way with turbulence and that the background ICM plasma
behaves more collisional \citep[see, e.g.,][]{Brunetti:2011ab}.
This is a more complex situation. It requires adequate calculations
where CRs are fully coupled with the turbulent evolution and 
in general leads to situations where the turbulent reacceleration
rate evolves with time.
We note however that the non-linear coupling between CRs and turbulence
does not change the functional form of the diffusion coefficient
in the particles momentum space, $D_{pp}$ (Sect.~3), but it generally 
increases its normalization implying a faster reacceleration \citep{Brunetti:2011ab}.
Thus the most important consequence would be that {\it less} turbulence 
is required to match a given reacceleration time $\tau_\mathrm{acc}$.

\noindent
We also note that the scaling 
that is expected in TTD reacceleration, $D_{pp} \propto p^2$, at least for
relativistic and ultrarelativistic particles, is fairly general for acceleration by 
large-scale turbulence. For example it is also common to betatron
acceleration or magnetic pumping
\citep[e.g.,][]{Melrose:1980,LeRoux:2015aa},
and to the situation where particles interact stochastically with
compression and rarefaction in compressive turbulence 
\citep[e.g.,][]{Ptuskin:1988aa,Cho:2006aa,Brunetti:2007aa}.
Since $D_{pp}$ (in combination with losses) determines the spectrum of 
accelerated particles, this implies that 
the spectral shapes of CR and consequently the ratio of gamma rays to
radio emission that are predicted {\it for a given reacceleration rate}, 
are not restricted to the particular use of the TTD mechanism.

Numerical simulations suggest that incompressible
turbulence is important in the ICM \citep{Miniati:2014aa,Vazza:2017aa}.
Turbulent reacceleration by large-scale incompressible turbulence 
in the ICM has been investigated in \citet{Brunetti:2016aa}.
In this case the reacceleration is not powered by compressions but 
it results from the interaction of particles with magnetic field
lines subject to turbulent diffusion.
In its most simple form, this mechanism leads to a functional form
of the diffusion coefficient that is equivalent to TTD, $D_{pp}
\propto p^2$, and - similarly to the case of collisional TTD and other mechanisms based on
compressive turbulence - the use
of this mechanism would simply lead to a different amount of turbulent
energy that is required to obtain a given acceleration rate.
In this case, however it is worth noting that 
the acceleration rate depends also on the
value of the beta-plasma and a situation of constant reacceleration rate 
in the cluster volume may appear less natural than in the case of the TTD.

\subsection{Limitations}

An obvious limitation of this paper is that 
we carried out calculations assuming spherical symmetry.
On one hand this allowed us a prompt comparison with the magnetic field values
estimated from Faraday RM
analysis and to use self consistently the values of the
thermal parameters of the ICM that are derived from X-ray observations under the same assumption.
On the other hand this simplification might have a potentially impact
on some of our conclusions.
The global ratio between radio and gamma-ray luminosities is the key
parameter in our analysis. In the case of hadronic models 
it is given by Eq.~\ref{ratioradiogamma} (for reacceleration it also
depends on the shapes of the spectrum of secondaries and CRp which are
modified by reacceleration and losses in a different way).
In the case $\eta_B \sim 0$ the ratio between gamma rays and radio does not depend on the geometry
of the system, on the other hand for steeper configurations of the magnetic 
field in the cluster the geometry may play a role.
Although the appearence of the spatial distribution of the gas of the Coma cluster
as projected on the plane of the sky is roughly symmetric (at least if
we exclude the sub-group in the South-East periphery), it is possible
that large deviations from the spherical symmetry occur in the
direction of the line of sight.
In the case of a spheroidal oblate (prolate), if we assume a
configuration of the magnetic field which declines with distance, 
the cluster gamma-ray to radio flux ratio would be smaller (larger)
than that estimated in the case of a spherical model.
These effects may become important for large departures from spherical
symmetry and in the cases of magnetic configurations with
large values of $\eta_B$ and small values of the central magnetic
fields. However, these magnetic configurations are clearly ruled out in the case of pure
hadronic models, since they produce very large gamma-ray fluxes, and
consequently these effects may have an impact only in the case of
reacceleration models. In this respect, future constrained 
numerical simulations \citep[see, e.g.,][]{Donnert:2010aa} will help in
evaluating in more detail the uncertainties of the predicted gamma-ray 
emission due to geometrical effects.

We also carried out calculations assuming that at each radius the
magnetic field strength and the particle number densities are homogeneus.
On the other hand spatial fluctuations of the magnetic field intensity
may affect the ratio between synchrotron and gamma rays. In general, since the
synchrotron emissivity scales non-linearly with magnetic field
intensity, the presence of magnetic fluctuations tend to increase the ratio
between radio and gamma rays.
Assuming viable configurations for the spatial variations of the magnetic
field, \citet{Brunetti:2012aa} have shown that the radio-synchrotron to gamma rays
ratio in the case of pure secondary models may increase up to a factor
2 with respect to homogeneus calculations, with the effect becoming 
progressively less important for larger magnetic fields (this is easy
to understand from Eq.~\ref{ratioradiogamma}). Since in the case of pure
hadronic models the new Fermi-LAT limits constrain very large values of the
magnetic field in the cluster it is very unlikely that the presence of 
magnetic field variations affect our conclusions.
On the other hand local variations of the magnetic field may affect the
case of reacceleration models. Under some circumstances, this would make the predicted 
gamma-ray emission slightly smaller than that calculated assuming
homogeneus conditions. However, a detailed evaluation of this effect requires the
adoption of specific probability distribution functions of the
magnetic fields values in the ICM combined with a quantitative analysis
of the spatial diffusion of electrons across magnetic field
fluctuations, which is beyond the aim of the present paper.

Another simplification in our calculations (and in
the formalism in Sect. 3) is that spatial diffusion of CRs
and propagation effects during the reacceleration phase
are not taken into account.
Calculations of turbulent reacceleration
are carried out for $\Delta T_\mathrm{acc}$ which 
in turns translates into
a resolution element where we have implicitly assumed
that the thermal, turbulent and magnetic field parameters are constant.
The size of this volume element is $L_{res} \sim 2 \sqrt{D \Delta
T_\mathrm{acc}}$, where $D$ is the spatial diffusion coefficient.
An upper limit to $L_{res}$ can be estimated by adopting a very 
optimistic view where particles
can travel undisturbed along field lines. In this case
diffusion is constrained by magnetic mirroring and by the tangling
scale of the magnetic
field by turbulent motions and the pitch-angle averaged diffusion
coefficient is $D \sim \langle
D_{\Vert \Vert} \rangle \sim \phi l_A c$, where $\phi \sim 0.1-0.2$ and
$l_A = M_A^{-3} L_0$ is the MHD scale (the scale where the eddy
velocity equal the Alfv\'en speed).
If we consider typical Alfv\'en Mach numbers and
turbulent driving scales in the ICM, about $M_A \sim 5$ and $L_0 \sim
100$ kpc, respectively, one
finds $D \sim 10^{31}$cm$^2$s$^{-1}$ and $L_{res} \leq 300
\sqrt{\Delta T_\mathrm{acc}/ Gyr}$ kpc.
Since we know that on scales $L_{res} > 200-300$ kpc in the ICM the physical
parameters that are used in Eqs.\ref{elettroni}--\ref{modes_kinetic}
change significantly, including the magnetic field strength, thermal
density and turbulent properties (assuming a typical injection scale of about few
$\times$100 kpc), we conclude that reacceleration periods
$\sim 1\,\mathrm{Gyr}$ are the largest ones that can be assumed for 
homogeneus calculations.

\section{Conclusions}

CRp are expected to be the most important non-thermal components in
galaxy clusters, still the energy budget that is associated with these
particles is difficult to predict due to our poor knowledge of the
micro-physics of the ICM, including the properties of particles
transport and acceleration in this plasma.
The combination of gamma rays and radio observations provides an
efficient way to constrain the energy budget of CRp in galaxy clusters and 
their role for the origin of cluster-scale radio emission.

\noindent
In this paper we have made an attempt to constrain the role of
relativistic protons for the origin of the prototype radio halo in 
the Coma cluster by combining radio and gamma-ray data.
More specifically we have assumed that secondary particles dominate
the synchrotron emission of the halo, model the properties of the radio
halo and explore the consequences for the gamma-ray emission of the cluster.
The radio modeling has been calculated as a function of the magnetic
field properties in the cluster and anchored to the brightness profile 
and flux of the radio halo at 350 MHz. 
In this way we derived the number density of CRp
with radial distance and calculated the resulting
gamma ray spectrum of the cluster.
We then used publicly available likelihood curves obtained from LAT observations 
to determine corresponding 95\% flux upper limits for our 
models and check whether they are consistent with one another.

\noindent
In order to have a prompt comparison with Faraday RM, 
we adopt a general formulation for the magnetic field strength and
scaling with radial distance in the cluster, $B(r) = B_0 (
n_{\mathrm{ICM}}(r)/n_{\mathrm{ICM}}(0))^{\eta_B}$, that is based on two parameters,
$B_0$ and $\eta_B$. As a consequence our study provides combined
constraints on $B_0$ and $\eta_B$.

\subsection{Conclusions on pure hadronic models}

The simplest case that we investigated is that where 
turbulence does not play a role.
This is the case of pure hadronic models.
In this simple scenario, for a given magnetic field configuration, the spectrum and spatial
distribution of CRp are then entirely constrained by the observed spectrum 
and brightness distribution of the radio halo.

Assuming the best-fit scaling from Faraday RM, where the magnetic field energy 
density in the cluster scales with the thermal density of the ICM
($\eta_B=0.5$), we derive $B \geq 21 \mu$G in 
the cluster core to be compared with $B \sim 4-5 \mu$G as inferred from 
Faraday RM.
The value of the minimum magnetic field decreases for sublinear
scalings ($\eta_B < 0.5$) but this also happens for the magnetic field values 
derived from RM. Specifically the ratio of the energy densities of the
magnetic field constrained by our analysis and that
measured from Faraday RM ranges from $> 14$ in the case
$\eta_B=0.2$ (i.e. the minimum $\eta_B$ allowed by RM) to $> 20$ in the case
$\eta_B =0.5$. Larger values of that ratio are derived for $\eta_B > 0.5$.
This improves very much the constraints derived by \citet{Brunetti:2012aa} 
where the ratio of energy densities were $> 3$ ($\eta_B =0.2$) and $> 4$ ($\eta_B=0.5$).

The constraints derived for the magnetic field 
have a broader impact on the dynamical role of the magnetic field in
the ICM. We have shown that, independently of $\eta_B$,
the magnetic field
pressure (volume averaged) at distance $R \geq 2.5\,r_{c}$, where most 
of the thermal and CRp energy budgets are contained, should be comparable 
to the thermal pressure ($\beta_\mathrm{pl} < 2-3$). In other words assuming 
an hadronic origin of the radio halo has the consequence that a
significant fraction of the energy of the Coma cluster must be in
the form of the magnetic field.
These findings confirm and significantly strengthen previous
conclusions and readily disfavor the hypothesis of
a pure hadronic origin of the radio halo.

\subsection{The reacceleration case}

Despite our results obtained for pure hadronic models,
we pointed out that CRp can still play a role for the origin of the
halo if these particles and their secondaries are reaccelerated by
turbulence.
In this case the tension with Fermi-LAT limits can be considerably 
alleviated because \lumi can be much smaller than in the case of a pure hadronic scenario.
In practice this results in the possibility that magnetic fields
that are much weaker than those in the pure hadronic case are allowed
by the gamma-ray limits. 

\noindent
The reacceleration scenario is however very complicated as it depends
on several parameters and requires extensive calculations to explore
a meaningful range of model parameters, which we have started to explore in this paper.

\noindent
We assume that the turbulent reacceleration rate is constant in the cluster volume.
Under the hypothesis that {\it only} CRp and their secondary particles are
present in the ICM the model parameters are the acceleration rate
$\tau_\mathrm{acc}^{-1}$, the duration of the reacceleration phase $\Delta
T_\mathrm{acc}$, and the magnetic field model $(B_0,\eta_B)$.
In this paper we have assumed a range of values: 
$\tau_\mathrm{acc} = 200-300$ Myr, $\Delta T_\mathrm{acc} = 350-1000$ Myr and studied   
two magnetic field models, with $\eta_B=0.5$ and 0.3, with 
$B_0$ as free parameter. 
Although we have explored a limited parameter range, in Sect.~6 we have 
shown that such a range is the relevant one to explain the radio halo.
This is the first systematic study of
the radio to gamma-ray spectrum from turbulent reacceleration 
of CRp and their secondaries assuming a range of model parameters, and 
in this respect it significantly extends
previous studies \citep{Brunetti:2007aa,Brunetti:2012aa,Pinzke:2017aa}.

\noindent
For central magnetic fields larger than a few $\mu$G 
all model configurations provide a sufficiently good representation of
the synchrotron spectrum of the Coma radio halo, including the observed
spectral steepening at high frequencies. However they generate
gamma-ray spectra that are significantly different.

\noindent
We find that the expected gamma-ray luminosity increases with 
increasing $\tau_\mathrm{acc}$ and with decreasing $\Delta T_\mathrm{acc}$, that is
essentially because in these models reacceleration boosts-up 
the spectrum of electrons at energies around GeV increasingly more than 
the spectrum of CRp in the case of faster reaccelerations 
and/or longer reacceleration periods.
As a matter of facts our study demonstrates that the combination of
radio and gamma-ray observations is a powerful tool to discriminate
between different situations and to constrain acceleration parameters
more broadly.

\noindent
Similarly to the case of pure hadronic models the comparison between
model expectations and Fermi-LAT limits allows to derive lower
limits to the values of the magnetic field in the cluster.
Assuming a reference case with $\tau_\mathrm{acc} = 260$ Myr and $\Delta
T_\mathrm{acc} = 720$ Myr, we obtained $B_0 > 4 \mu$G in the case $\eta_B = 0.5$
and $B_0 > 2.5 \mu$G for $\eta_B =0.3$. 
These limits are much weaker
than those derived under the assumption of pure hadronic models and 
in fact are compatible with values from the analysis of Faraday RM.
Higher (lower) values of the 
mimimum magnetic field are obtained for larger (smaller) $\tau_\mathrm{acc}$
and for smaller (larger) $\Delta T_\mathrm{acc}$. 
In general the magnetic field values constrained by our analysis are 
not in tension with values inferred from RM. However the model
configurations 
combining lower reacceleration rates and shorter acceleration periods
require large magnetic fields that can be in tension with Faraday RM. 
For example, in the reference case $\eta_B=0.5$, 
model configurations with $\Delta T_\mathrm{acc} \leq 400-450$ Myr require magnetic 
fields that are larger than those from 
Faraday RM across the entire range of values 
of reacceleration rates that is explored in our calculations.
Flatter spatial distributions of the magnetic field, $\eta_B
< 0.5$, generate smaller gamma-ray fluxes allowing situations with smaller
magnetic fields also in the configurations where reacceleration
periods are short. However the tension with RM in the case of short
reacceleration periods is not 
significantly reduced also for $\eta_B < 0.5$ because in this case also the 
values of the magnetic fields that are constrained by RM are smaller.

\noindent
Interestingly
we have shown that a non detection of the Coma cluster
using 10-15 yrs of Fermi-LAT data would imply a 
magnetic field in the cluster that is in clear
tension with Faraday RM.
This also open to the possibility of future detection of the Coma
cluster in the gamma rays, provided that CRp play a role for the origin
of the radio halo. We concluded that
at least under a number of assumptions (namely quasi-homogeneus magnetic fields and
quasi-spherical symmetry) there is a
chance to obtain a signal in excess of 3$\sigma$ for
models that assume magnetic field configurations compatible with
Faraday RM.

\subsection{Constraints on the energy content and spatial distribution 
of CRp}

The Fermi-LAT limits constrain the maximum level of the
energy budget that can be associated to CRp. 
The budget that is allowed increases for
flatter spatial distributions of the CRp and this is
because less gamma rays are
generated due to the fact that the density of thermal targets in the
cluster declines with distance.
Several papers suggest that CRp distribute more broadly than the
thermal ICM \citep[e.g.,][]{Vazza:2014aa,Wiener:2013aa,2011A&A...527A..99E}, 
still our poor knowledge of CRp dynamics in the
ICM does not allow solid conclusions.

\noindent
If radio halos have hadronic origin, 
radio observations set combined constraints on the
spatial distrbution of CRp and magnetic field in the clusters.
The broad brightness profile of the Coma radio halo implies
that the underlying spatial distribution of the CRp is very broad as well.
We find that the ratio of CRp and thermal energy densities should increase
significantly with
radius even in the case of a magnetic field profile that is much
flatter than that of the thermal gas (Figs. 4 and 10) with the
consequence that most of the energy of CRp should be stored in the
external regions, $r \geq 2.5-3 r_c$; a situation
that is challenging to understand \citep[e.g.,][]{Zandanel:2014ab}.

\noindent
The combination of radio and gamma-ray data allows to put combined
constraints on $B_0$ and $\eta_B$ and to break the degeneracy between
the spatial distributions of magnetic field and CRp.
Importantly, we have shown that the current Fermi-LAT observations limit the level of
the energy budget of CRp to $\leq 10\%$ of the thermal ICM in a way that
does not depend very much (within a factor 2) on the specific model
of the halo (pure hadronic or reacceleration) and on the value of $\eta_B$.
However, such a level of CRp energy content, which is increasing at larger radii, 
could undermine clusters' hydrostatic mass estimates where
non-thermal components, such as turbulence, CRs, and magnetic fields, are 
usually ignored (e.g., \citealp{2008MNRAS.385.2243A}). In this sense, despite the great
steps forward obtained thanks to gamma-ray observations, we are 
still far from sub-percent-level constrains or from a detailed
knowledge of CRs in clusters.

\noindent
Of course it is possible that the 
energy budget of CRp in the Coma cluster is significantly below this upper limit,
simply because the limit is obtained under the 
hypothesis that secondary particles dominate the synchrotron emission
up to large distances from the center. 
Therefore, it could be that the 
CRp follow a steeper distribution and that the external regions of the
halo are powered by the reacceleration of primary electrons.
In this case the energy budget of CRp as constrained by the 
Fermi-LAT limits would be much smaller.

\section*{Acknowledgements}
GB acknowledges partial supported by INAF under grant PRIN-INAF 2014.
FZ acknowledges the support of the Netherlands Organization for Scientific 
Research (NWO) through a Veni grant. SZ acknowledges the hospitality of 
GRAPPA where parts of this work were carried out.

%\bibliographystyle{mnras}
%\bibliography{references2}

%%%%%%%%%%%%%%%%%%%%%%%%%%%%%%%%%%%%%%%%%%%%%%%%%%

%%%%%%%%%%%%%%%%% APPENDICES %%%%%%%%%%%%%%%%%%%%%

\appendix

\section{$D_{pp}$ Formula}

In this Appendix we derive a general 
formula for the diffusion coefficient in the momentum
space due to TTD with fast modes in the ICM
that is valid also in the non-relativistic regime.

\noindent
The resonant condition for TTD is:
\begin{equation}
\omega = k_{\Vert}v_{\Vert}
\label{resonance}
\end{equation}
\noindent
where $\omega = c_{s} k$ is the frequency of fast modes and
$v_{\Vert}=\mu v$ and $k_{\Vert}=\eta k$
are the parallel (projected along the magnetic field)
speed of the particles and the wave--number, respectively.
This resonance changes only the
component of the particle momentum parallel to the
seed magnetic field and this would cause an increasing degree of
anisotropy of the particle distribution leading to a less and less
efficient process with time. However it is very likely that isotropization of 
particle momenta during acceleration is
preserved by some mechanism \citep[e.g.,][for
discussions]{Schlickeiser:1998aa,Brunetti:2007aa},
and for this reason in our calculations we will assume a isotropic distribution
of particles.

In order to derive the acceleration coefficient we follow
\citet{Brunetti:2007aa} and use an approach based on the {\it detailed balancing}.
\noindent
The starting point is the collisionless damping rate of the waves due
to TTD, that measures the rate of dissipation of turbulent energy into
particles via TTD.
The damping rate of fast modes due to TTD with $\alpha$-species
particles in the ICM was derived by \citet{Brunetti:2007aa}:
\begin{equation}
\Gamma_{\alpha} =
- {{\pi^2}\over{8}}
{{|B_k|^2}\over{B_0^2}} {{c_s k}\over{W}} {{\mu (1-\mu^2)}\over{|\mu|}}
{{ {\cal H}(1-\big| {{c_s}\over{c \mu}} \big| ) }\over
{\sqrt{1 - ( {{c_s }\over{c \mu }} )^2 }}} {\cal I}_{\alpha}
\label{damping_perp}
\end{equation}
\noindent
where $\mu$ is the wave pitch-angle (i.e., $k_{\Vert}/k$) and 
\begin{equation}
{\cal I}_{\alpha} = {1 \over{m_{\alpha} }} \int_0^{\infty} dp_{\perp}
{{p_{\perp}^5 }\over{ \sqrt{ 1 + ( {{p_{\perp}}\over{m_{\alpha} c}}
)^2 } }}
\left(
{{\partial f_{\alpha} }\over{ \partial p_{\Vert} }}
\right)_{p_{\Vert , res}} \, .
\label{integral}
\end{equation}

\noindent
The integral (Eq.~\ref{integral}) has to be evaluated at the resonant
momentum (i.e., using Eq.~\ref{resonance}), specifically under the condition that the parallel momentum is:
\begin{equation}
p_{\Vert , res} = m_{\alpha} c \left( {{c_s}\over{c \mu}}
\right)
\left(
{{ 1 + ( {{p_{\perp}}\over{m_{\alpha} c}} )^2 }\over{
1 - ( {{c_s}\over{c \mu}} )^2 }} \right)^{1 \over 2} \, .
\label{pres}
\end{equation}

\noindent
Under the assumption of isotropic distribution of particles,
the next step is to re-write Eqs.~\ref{damping_perp}--\ref{integral}
using particle momentum $p=\sqrt{p_{\Vert}^2+p_{\perp}^2}$.
Eq.~\ref{pres} gives:
\begin{equation}
d p_{\perp} =
{{ 1 - ( {{c_s}\over{c \mu}} )^2 }\over{
\sqrt{ 1 - ( {{c_s}\over{c \mu}} )^2/\beta^2 }}}
d p_{p_{\Vert, res}}
\end{equation}
\noindent
and the damping rate is:
\begin{equation}
\Gamma_{\alpha}(k,\mu)=
\tilde{ \Gamma}_{\alpha}  \int_{p_{m}}
dp p^4 \big[ 1 - 
( {{c_s}\over{c \mu}} )^2/\beta^2 \big]^2
\left( {{\partial f}\over{\partial p}} \right)
{\cal H}(1 - {{c_s}\over{c \mu \beta}})
\label{damping_p}
\end{equation}
\noindent
where 
\begin{equation}
\tilde{\Gamma}_{\alpha} = 
{{\pi^2}\over{8}} k 
{{|B_k|^2}\over{B_0^2}} {{c_s^2}\over{W}}
{{\mu (1-\mu^2)}\over{|\mu|}} \,\,\, ,
\label{tilde}
\end{equation}
\noindent
the Heaviside function in Eq.~\ref{damping_p} accounts for the
condition $p_{\Vert, res} < p$, and the minimum momentum in the
integral in Eq.~\ref{damping_p} is:
\begin{equation}
p_{m}= p_{\Vert}(p_{\perp} \rightarrow 0)=
{{m_{\alpha} c_s }\over{
1 - ( {{c_s }\over{c \mu }} )^2 }} {1 \over{\mu}} \, .
\end{equation}

The {\it detailed balance} approach accounts for the fact
that the rate of energy that
is extracted from the turbulence through collisionless damping with 
particles species $\alpha$, $\int d^3k W \Gamma $,
is channelled into the same particles:
\begin{equation}
\int d^3p E_{\alpha} {{\partial f }\over{\partial t}}
=
\int d^3k \Gamma(k,\mu) W(k) \, .
\label{balance}
\end{equation}

\noindent
Assuming isotropy, the time derivative of the particle
phase-space distribution function is directly connected with the
particle momentum diffusion coefficient via a Fokker-Planck equation:
\begin{equation}
{{\partial f}\over{\partial t}}=
{{ 1 }\over{p^2}}
{{\partial }\over{\partial p}} \big[
p^2 D_{pp} {{\partial f}\over{\partial p}}
\big]
\label{FPiso}
\end{equation}
\noindent
where $D_{pp}$ is the momentum diffusion coefficient due to the
same mechanism that drains energy from the turbulence into particles
at the rate $\int d^3k W \Gamma$.

\noindent
Combining Eqs.\ref{balance}--\ref{FPiso} allows to connect directly the
momentum diffusion coefficient with the damping coefficient
of turbulence:
\begin{equation}
\int_{p_m} d^3p {{E_{\alpha}}\over{p^2}}
{{\partial }\over{\partial p}} \big[
p^2 D_{pp} {{\partial f}\over{\partial p}}
\big]
=
\int d^3k \Gamma(k,\mu) W(k) \, .
\label{balance2}
\end{equation}

\noindent
The left-hand side of Eq.~\ref{balance2} can be calculated via partial
integration that gives:
\begin{equation}
\int d^3p {{E_{\alpha}}\over{p^2}}
{{\partial }\over{\partial p}} \big[
p^2 D_{pp} {{\partial f}\over{\partial p}}
\big]
=
- {{4 \pi}\over{m_{\alpha}}}
\int p^3 D_{pp} \beta
{{\partial f }\over{\partial p}} dp \, .
\label{partial}
\end{equation}

\noindent
As a consequence the combination of Eqs.\ref{balance2}--\ref{partial} and
\ref{damping_p}--\ref{tilde} allows to derive 
the expression for the momentum diffusion coefficient:
\begin{eqnarray}
D_{pp} \simeq
{{\pi^2}\over{2 c}}
{{c_s^2}\over{B_0^2}}
{{p^2}\over{\beta}}
\int_{{c_s}\over{\beta c}}^1
{\cal H}(1 - {{c_s}\over{\beta c}})
{{1 - \mu^2}\over{\mu}} d \mu \nonumber\\
\left[ 1 - \big( {{c_s}\over{\mu \beta c}} \big)^2
\right]^2 
\int dk k W_B
\label{dppfinal}
\end{eqnarray}
\noindent
where in obtaning Eq.~\ref{dppfinal} we have introduced the 
isotropic spectrum (energy density per wavenumber) of
magnetic field fluctuations associated to fast modes using:
\begin{equation}
W(k) = {{W_B}\over{4 \pi k^2 }}
\left(
{{16 \pi W}\over{|B_k|^2}}
\right)
\end{equation}
\noindent
where $16 \pi W / |B_k|^2 \sim {\cal O}(1)$ \citep{Brunetti:2007aa}.
Note that in the limit
$p \gg m c$ Eq.~\ref{dppnew} is Eq.~(40) in \citet{Brunetti:2007aa}.

\begin{figure*}
\includegraphics[width=0.9\textwidth]{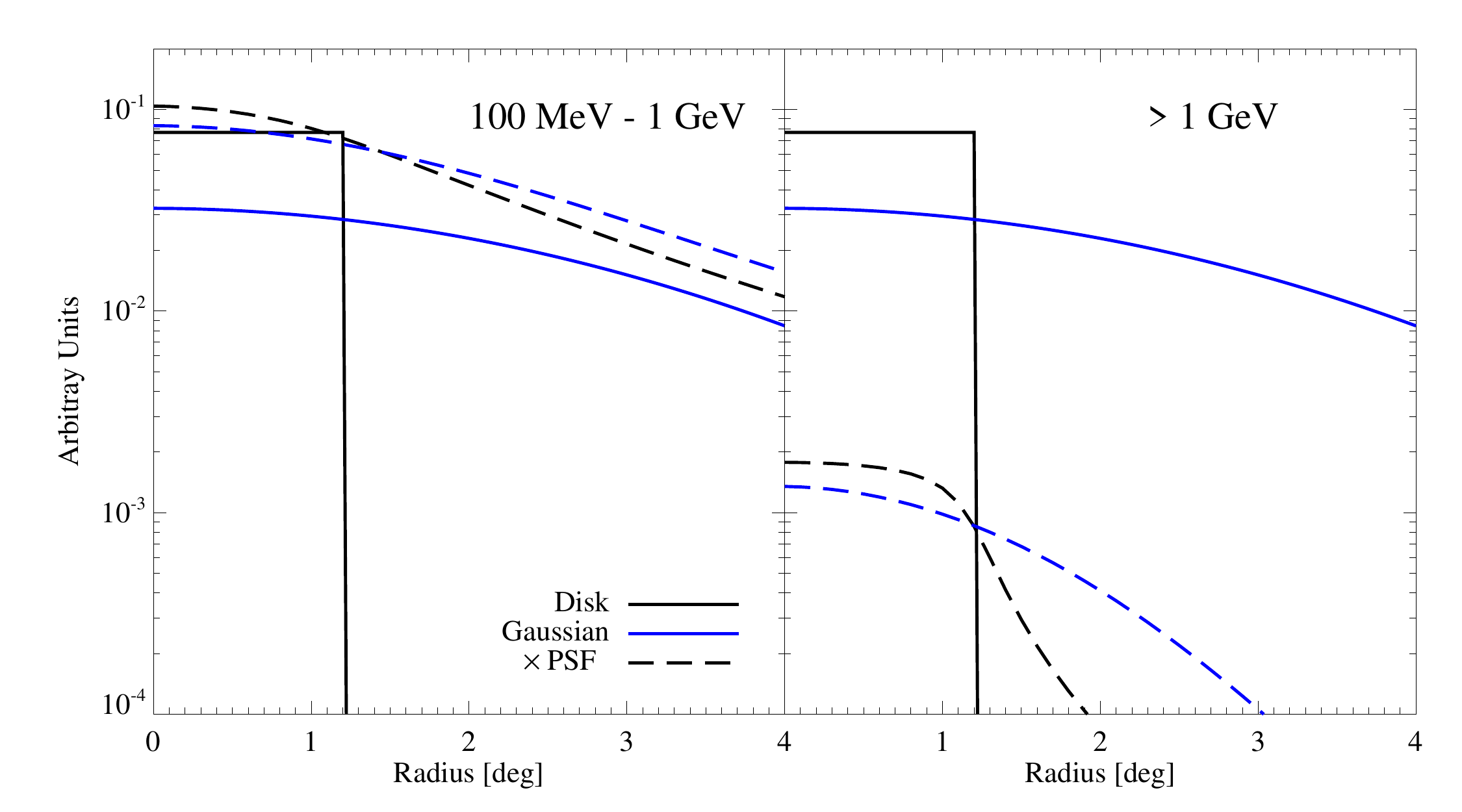}
\caption{Radial surface brightness profile (arbitrary units) 
assuming a disk (black-solid line) and a Gaussian emission profile (blue-solid line), 
folded with the PSF for $P8R2\_SOURCE\_V6$ IRF (dashed lines) for low energies (100 MeV - 1 GeV) 
and at higher energies (> 1 GeV). Aside from a different normalization (which however is a free 
parameter in the likelihood procedure), at low energies, both models cannot be effectively distinguished from one another. }
\label{fig:Lande}
\end{figure*}

\section{Validation of Analysis\label{sec:validation}}

In this section we discuss the validity of our assumption that given the large PSF at low energies we can express 
any radially symmetric profile as an isotropic disk of unknown radius $R$. 

\noindent
Using the conversion of an incoming gamma-ray into an
electron-positron pair, Fermi-LAT employs a silicon tracker
interleaved with tungsten plates to promote pair conversion
\citep{Atwood:2009aa}. As a
result, the angular resolution, or point-spread function (PSF) becomes
a steeply varying function with energy (and incidence angle). As
previoulsy shown in \citet{Lande:2012aa},
a model selection between a 2D Gaussian-like template for a
source of size $\sigma$:
\begin{equation}
I_{\mathrm{Gaussian}}(x,y)=\frac{1}{2\pi\sigma^{2}}\times\exp\left(-(x^2
+ y^2)/2\sigma^2\right)
\end{equation}
and an isotropic disk-like template of similar size,
cannot easily be made despite their analytical shapes being
significantly different:
\begin{equation}
I_{\mathrm{Disk}}(x,y)=\begin{cases}\frac{1}{\pi\sigma^2}\quad
&x^2+y^2 \leq \sigma^2\\0\quad &x^2+y^2 > \sigma^2. \end{cases}
\end{equation}

While the underlying reconstruction software has received a major overhaul from \cite{Lande:2012aa} 
to the data in \cite{Ackermann:2016aa}, we expect that for faint 
and largely extended gamma-ray emitters as it is expected from the diffuse gamma-ray emission 
from  CR interactions in the ICM, the results of \citet{Lande:2012aa} still apply.\footnote{Note that \citet{Lande:2012aa} 
considered Galactic supernova remnants as target for their study, which compared to cluster-scales are much smaller.} 

Part of the output of {\tt{gtsrcmaps}} is a 2D model map in which the input model is folded with the IRFs 
evaluated in the energy bin $E_j$. We sum up each of these maps above $E_{\mathrm{min}}$ weighted by a power-law 
spectrum with index -2.0 and extract the radial profile from the center of the map. 
In Fig.~\ref{fig:Lande} we show the resultant radial profiles for an isotropic disk with radius
as the virial radius of the Coma cluster $\theta_\mathrm{Coma}$ compared with a Gaussian with $\sigma=\theta_\mathrm{Coma}$ above 100 MeV (left) and 1 GeV (right). 
A visual comparison between this figure and Fig.~2 in \citet{Lande:2012aa} supports our assumption, especially 
considering the quoted systematic uncertainties in \citet{Ackermann:2016aa}, which are dominated by the 
details of the Galactic foreground modeling. 

Finally, we consider the effect of the existing grid of likelihood curves of varying disks by calculating 
modified flux upper limits for disk radii which do not match according to our comparison operators. 
Specifically, if for a given model $M_{i}$ we find a matching radius $r_{i}$, we calculated the resulting limits 
assuming that $r_{i\pm1}$ are chosen instead. We find that the limits change less than $\pm15\%$ with respect to 
the optimal value for $r_{i}$ (irrespective of the chosen comparison operator). Note that this value is well 
below the overall systematic uncertainties quoted in \citet{Ackermann:2016aa}, where bin-wise uncertainties 
are given as $<22\%$ for energy bins with $E>300\,\mathrm{MeV}$ and $<54\%$ towards the lower energies and 
integral limits are quoted with $<21\%$ for a hard spectral index ($\Gamma=1.6$) and $<42\%$ 
for a soft spectral index ($\Gamma=2.6$), respectively.

%%%%%%%%%%%%%%%%%%%%%%%%%%%%%%%%%%%%%%%%%%%%%%%%%%

% Don't change these lines
\bsp	% typesetting comment
\label{lastpage}
\end{document}